\documentclass[AMA,STIX1COL]{WileyNJD-v2}

\usepackage{enumitem}
\usepackage[capitalise]{cleveref}
\usepackage{csquotes}
\usepackage{threeparttable}
\usepackage{colortbl}
\usepackage{subcaption}

\usepackage{booktabs}
\definecolor{Gray}{RGB}{100,100,100}
\definecolor{naturelight}{RGB}{248,231,215}
\definecolor{naturedark}{RGB}{204,139,136}
\definecolor{myblue}{RGB}{0,100,200}
\definecolor{myred}{RGB}{204,102,0}

\usepackage{siunitx}
\DeclareSIUnit\mmHg{mmHg}
\usepackage{nomencl}
\makenomenclature
\usepackage{etoolbox}
\renewcommand\nomgroup[1]{%
  \item[\bfseries
  \ifstrequal{#1}{abbr}{Abbreviations}{%
  \ifstrequal{#1}{aw}{Parameters of airway elements}{%
  \ifstrequal{#1}{tu}{Parameters of terminal units}{%
  \ifstrequal{#1}{input}{Input parameters for RD parametrization}{%
  \ifstrequal{#1}{ppl}{Fitting variables for patient-specific pleural pressure}{%
  \ifstrequal{#1}{vent}{Ventilation constants}{}}}}}}%
]}

\hypersetup{
	colorlinks,
	linkcolor=myred,
	citecolor=myblue,
	urlcolor=myblue
}

\articletype{Applied Research}%

\received{}
\revised{}
\accepted{}

\begin{document}

\title{Pressure- and time-dependent alveolar recruitment/derecruitment in a spatially resolved patient-specific computational model for injured human lungs}

\author[1]{Carolin M. Geitner*}

\author[1]{Lea J. K\"oglmeier}

\author[2]{In\'{e}z Frerichs}

\author[3]{Patrick Langguth}

\author[2]{Matthias Lindner}

\author[2]{Dirk Sch\"adler}

\author[2]{Norbert Weiler}

\author[2]{Tobias Becher}

\author[1]{Wolfgang A. Wall}

\authormark{GEITNER \textsc{et al}}

\address[1]{\orgdiv{Institute for Computational Mechanics}, \orgname{Technical University of Munich}, \orgaddress{\state{Garching b. Muenchen}, \country{Germany}}}

\address[2]{\orgdiv{Department of Anesthesiology and Intensive Care Medicine}, \orgname{University Medical Center Schleswig-Holstein, Campus Kiel}, \orgaddress{\state{Kiel}, \country{Germany}}}

\address[3]{\orgdiv{Department of Radiology and Neuroradiology}, \orgname{University Medical Center Schleswig-Holstein, Campus Kiel}, \orgaddress{\state{Kiel}, \country{Germany}}}

\corres{*Carolin M. Geitner, Institute for Computational Mechanics, Technical University of Munich, Boltzmannstrasse 15, 85748 Garching b. Muenchen, Germany. \\ \email{carolin.geitner@tum.de}}

\abstract[Summary]{%

We present a novel \textcolor{black}{computational model for} the dynamics of alveolar recruitment/derecruitment~(RD), which reproduces the underlying characteristics typically observed in injured lungs. The basic idea is a pressure- and time-dependent variation of the stress-free reference volume in reduced dimensional viscoelastic elements representing the acinar tissue. We choose a variable reference volume triggered by critical opening and closing pressures in a time-dependent manner from a straightforward mechanical point of view. In the case of (partially and progressively) collapsing alveolar structures, the volume available for expansion during breathing reduces and vice versa, eventually enabling consideration of alveolar collapse and reopening in our model. We further introduce a method \textcolor{black}{for patient-specific determination of} the underlying critical parameters of the new alveolar RD dynamics when integrated into the tissue elements, referred to as terminal units, of a spatially resolved physics-based lung model that simulates the human respiratory system in an anatomically correct manner. Relevant patient-specific parameters of the terminal units are herein determined based on medical image data and the macromechanical behavior of the lung during artificial ventilation. We test the whole modeling approach for a real-life scenario by applying it to the clinical data of a mechanically ventilated patient. The generated lung model is capable \textcolor{black}{of reproducing} clinical measurements such as tidal volume and pleural pressure during various ventilation maneuvers. We conclude that this new model is an important step \textcolor{black}{toward} personalized treatment of ARDS patients by considering potentially harmful mechanisms~---~such as cyclic RD and overdistension~---~and might help \textcolor{black}{in the development of} relevant protective ventilation strategies to reduce ventilator-induced lung injury (VILI).}


\maketitle

\nomenclature[abbr]{$\mathrm{HU}$}{Hounsfield Unit}

\nomenclature[aw]{$C$}{Capacitance of an airway element}
\nomenclature[aw]{$I$}{Inductance of an airway element}
\nomenclature[aw]{$R$}{Resistance of an airway element}
\nomenclature[aw]{$P_{\mathrm{in}}$}{Pressure at the inlet of an airway element}
\nomenclature[aw]{$P_{\mathrm{out}}$}{Pressure at the outlet of an airway element}
\nomenclature[aw]{$\widetilde{P}_{\mathrm{ext}}$}{External pressure of an airway element}
\nomenclature[aw]{$Q_{\mathrm{out}}$}{Outflow of an airway element}
\nomenclature[aw]{$Q_{\mathrm{in}}$}{Inflow of an airway element}

\nomenclature[tu]{$N_\mathrm{ad}$}{Number of alveolar ducts in a terminal unit}
\nomenclature[tu]{$V_{\mathrm{PEEP}}$}{Gas volume of a terminal unit at PEEP}
\nomenclature[tu]{$V_{\mathrm{water}}$}{Water volume of a terminal unit}
\nomenclature[tu]{$V_{\mathrm{tissue}}$}{Tissue volume of a terminal unit}
\nomenclature[tu]{$V_{\mathrm{edema}}$}{Edema volume of a terminal unit}
\nomenclature[tu]{$Q$}{Gas flow into a terminal unit}
\nomenclature[tu]{$P$}{Difference pressure across a terminal unit}
\nomenclature[tu]{$P_{\mathrm{alv}}$}{Alveolar pressure inside a terminal unit}
\nomenclature[tu]{$B$}{Dashpot equal for all terminal units}
\nomenclature[tu]{$B_\mathrm{a}$}{Dashpot equal for all terminal units}
\nomenclature[tu]{$P_\mathrm{E1}$}{Non-linear stiffness of a terminal unit}
\nomenclature[tu]{$E_\mathrm{2}$}{Linear stiffness equal for all terminal units}
\nomenclature[tu]{$V$}{Current gas volume of a terminal unit}
\nomenclature[tu]{$V_{\mathrm{0}} \left( t \right)$}{Stress-free reference gas volume of $V$ of a terminal unit at time $t$}
\nomenclature[tu]{$V_{\mathrm{0,init}}$}{Initial reference gas volume of a terminal unit}
\nomenclature[input]{$\kappa$}{Stiffness parameter in $P_\mathrm{E1}$ equal for all terminal units}
\nomenclature[input]{$\beta$}{Shape determining parameter in $P_\mathrm{E1}$ equal for all terminal units}
\nomenclature[tu]{$V_{\mathrm{0,targ}} \left( P \right)$}{Target reference gas volume of a terminal unit}
\nomenclature[tu]{$V_{\mathrm{0,min}}$}{Minimal reference gas volume of a terminal unit}
\nomenclature[tu]{$V_{\mathrm{0,max}}$}{Maximal reference gas volume of a terminal unit}
\nomenclature[tu]{$P_{\mathrm{crit,min}}$}{Minimal critical pressure of a terminal unit}
\nomenclature[tu]{$P_{\mathrm{crit,max}}$}{Maximal critical pressure of a terminal unit}
\nomenclature[tu]{$\tau$}{RD time constant of a terminal unit}
\nomenclature[tu]{$\tau_\mathrm{insp}$}{Inspiratory RD time constant of a terminal unit}
\nomenclature[tu]{$\tau_\mathrm{exp}$}{Expiratory RD time constant of a terminal unit}
\nomenclature[tu]{$P_{\mathrm{cl,crit,min}}$}{Minimal critical pressure of the closing path of a terminal unit}
\nomenclature[tu]{$P_{\mathrm{cl,crit,max}}$}{Maximal critical pressure of the closing path of a terminal unit}
\nomenclature[tu]{$P_{\mathrm{op,crit,min}}$}{Minimal critical pressure of the opening path of a terminal unit}
\nomenclature[tu]{$P_{\mathrm{op,crit,max}}$}{Maximal critical pressure of the opening path of a terminal unit}
\nomenclature[tu]{$h$}{Ventral-to-dorsal height of a terminal unit in the lung}
\nomenclature[tu]{$V_{\mathrm{0,PEEP}}$}{Reference gas volume of a terminal unit at PEEP}
\nomenclature[tu]{$V_{\mathrm{0,targ,ee}}$}{Target reference gas volume of a terminal unit at end-expiration during normal ventilation}
\nomenclature[tu]{$V_{\mathrm{0,targ,ei}}$}{Target reference gas volume of a terminal unit at end-inspiration during normal ventilation}
\nomenclature[tu]{$V_{\mathrm{0}}(P,t)$}{Current reference gas volume of a terminal unit}
\nomenclature[tu]{$V_{\mathrm{0},n}$}{Current reference gas volume of a terminal unit at end-inspiration or end-expiration}
\nomenclature[tu]{$V_{\mathrm{0,insp}}$}{Current reference gas volume of a terminal unit at end-inspiration}
\nomenclature[tu]{$V_{\mathrm{0,exp}}$}{Current reference gas volume of a terminal unit at end-expiration}
\nomenclature[tu]{$c_{\mathrm{I}}$}{Constant}
\nomenclature[tu]{$c_{\mathrm{E}}$}{Constant}

\nomenclature[input]{$\Delta P_{\mathrm{max-min}}$}{Pressure difference between minimal and maximal critical pressures}
\nomenclature[input]{$\Delta P_{\mathrm{op-cl}}$}{Pressure difference between opening and closing path}
\nomenclature[input]{$\epsilon_\mathrm{V_\mathrm{0}}$}{Tolerance factor}
\nomenclature[input]{$T$}{Time constant}
\nomenclature[input]{$\mu_{\mathrm{cl}}$}{Mean of $P_{\mathrm{cl,crit,max}}$ for normally ventilated terminal units}
\nomenclature[input]{$\sigma_{\mathrm{cl}}$}{Standard deviation of $P_{\mathrm{cl,crit,max}}$ for normally ventilated terminal units}
\nomenclature[input]{$\Delta P_{\mathrm{op}}$}{Pressure difference}
\nomenclature[input]{$\sigma_{\mathrm{op}}$}{Standard deviation of $P_{\mathrm{op,crit,min}}$ for collapsed terminal units}
\nomenclature[input]{$k_{\mathrm{coll}}$}{Volume collapse factor}
\nomenclature[input]{$k_\mathrm{edema}$}{Factor for edema volume}

\nomenclature[ppl]{$P_{\mathrm{pl}}^\mathrm{vol}$}{Volume-dependent pleural pressure component of $P_{\mathrm{pl}}$}
\nomenclature[ppl]{$P_{\mathrm{pl}}^\mathrm{weight}$}{Weight-dependent pleural pressure component of $P_{\mathrm{pl}}$ on a terminal unit depending on its $h$}
\nomenclature[ppl]{$a_\mathrm{v}$, $b_\mathrm{v}$, $c_\mathrm{v}$, and $d_\mathrm{v}$}{Patient-specific parameters of $P_{\mathrm{pl}}^\mathrm{vol}$} 
\nomenclature[ppl]{$V_{\mathrm{frac}}$}{Volume share of the gas volume increase in all terminal units}
\nomenclature[ppl]{$V_{\mathrm{tot}}$}{Gas volume of all terminal units}
\nomenclature[ppl]{$V_{\mathrm{tot,PEEP}}$}{Gas volume of all terminal units at PEEP}
\nomenclature[ppl]{$V_{\mathrm{tot,max}}$}{Maximal gas volume of all terminal units during quasi-static inflation maneuver}
\nomenclature[ppl]{$a_{\mathrm{w}}$, $b_{\mathrm{w}}$, $c_{\mathrm{w}}$, $d_{\mathrm{w}}$, and $e_{\mathrm{w}}$}{Patient-specific parameters of $P_{\mathrm{pl}}^\mathrm{weight}$}
\nomenclature[ppl]{$h_{\mathrm{balloon}}$}{Height of the esophageal balloon}
\nomenclature[ppl]{$P_{\mathrm{pl}}$}{External pressure depending on $V_{\mathrm{tot}}$ and $h$ of a terminal unit}

\nomenclature[vent]{$\mathrm{PEEP}$}{Positive end-expiratory pressure}
\nomenclature[vent]{$P_{\mathrm{tp,PEEP}}$}{Transpulmonary pressure of a terminal unit at PEEP (applied during the CT recording)}
\nomenclature[vent]{$P_{\mathrm{tp,endinsp}}$}{Transpulmonary pressure of a terminal unit at end-inspiration}
\nomenclature[vent]{$P_{\mathrm{endinsp}}$}{Airway pressure of a terminal unit at end-inspiration}
\nomenclature[vent]{$t_{\mathrm{insp}}$}{Time of inspiration in a normal breath cycle}
\nomenclature[vent]{$t_{\mathrm{exp}}$}{Time of expiration in a normal breath cycle}
\nomenclature[vent]{$P_{\mathrm{tp,qs,start}}$}{Transpulmonary pressure of a terminal unit at the start of the quasi-static inflation maneuver}
\nomenclature[vent]{$P_{\mathrm{tp,qs,max}}$}{Transpulmonary pressure of a terminal unit at peak inspiration of the quasi-static inflation maneuver}
\nomenclature[vent]{$P_{\mathrm{tp,qs,end}}$}{Transpulmonary pressure of a terminal unit at end-expiration of the quasi-static inflation maneuver}
\nomenclature[vent]{$t_{\mathrm{insp,qs}}$}{Time of inspiration in the quasi-static inflation maneuver}
\nomenclature[vent]{$t_{\mathrm{exp,qs}}$}{Time of expiration in the quasi-static inflation maneuver}

\section{Introduction}
\label{sec:1}

In patients with acute respiratory distress syndrome (ARDS), mechanical ventilation is a potentially life-saving treatment. Nevertheless, improperly adjusted ventilator settings can lead to ventilator-associated lung injury~(VILI)~\cite{Dreyfuss1998,Gajic2004,Slutsky2013}. Despite intense research in this field, which has among other things revealed two major contributors to VILI~---~i.e.,~overdistension~(volutrauma) and cyclic (re-)opening~(atelectrauma) of lung structures~\cite{Slutsky2013,Ramcharran2022}~---~, the development of protective ventilation strategies stagnated over the last two decades~\cite{Villar2016,Nieman2020}. 

The main reason for this halt is the inaccessibility of insights into regional mechanics of patients' lungs and damaging processes occurring during ventilation~\cite{Kollisch-Singule2018}. The situation is particularly exacerbated by the unique \textcolor{black}{heterogeneous} pathology of every diseased organ causing irregular and unpredictable tissue straining, tidal recruitment and distribution of ventilation~\cite{Kollisch-Singule2018,Cressoni2014,Kollisch-Singule2020,Fardin2021}. Local stress raisers, \textcolor{black}{that is, sites of high and thus harmful stress, which presumably occur in the presence of lung inhomogeneity and dysfunction~\cite{Gattinoni2012,Bilek2003} and were identified as the origin of injuries~\cite{Ramcharran2022,Fardin2021,Retamal2014},} cannot be determined in routine clinical practice, i.e., the impact of an applied ventilation protocol on the regional microscale~---~may it be beneficial or harmful~---~is hardly assessable, in particular for a specific patient~\cite{Nieman2020,Kollisch-Singule2018,Cressoni2014,Kollisch-Singule2020}. As a result, the current clinical practice of generic ventilation management and ventilator adjustment at the bedside~\cite{Ranieri2012} reaches its limits. An individual \textcolor{black}{heterogeneous} lung injury needs an individual therapy management~\cite{Gattinoni2017a,Reddy2020}. However, such management is only possible by achieving a locally deeper, patient-specific understanding of ARDS lung mechanics~\cite{Kollisch-Singule2018,Gattinoni2017a}.


Computational models of the human lung are promising tools for serving this purpose non-invasively. A broad range of modelling approaches already exist~\cite{Roth2017e,Neelakantan2022b}. By virtue of their high computational efficiency, reduced dimensional models in particular are feasible for use in clinical application.

In the context of VILI, capturing both overdistension~(OD) and cyclic recruitment and derecruitment~(RD) is crucial in enabling the application of these models in order to minimize injury~\cite{Ramcharran2022}. In general, the existing lung models often only concentrate on one of the two mechanisms~\cite{Salmon1981,Frazer2004,Sundaresan2009b,Ma2010,Smith2013,Ryans2016,Broche2017,Roth2017a}, lack regional resolution~\cite{Salmon1981,Frazer2004,Ma2010,Smith2013,Ryans2016,Hamlington2016,Amini2017a,Mellenthin2019}, or do not~---~or, at best, only statistically~---~consider the patient-specific pathology/heterogeneity~\cite{Salmon1981,Sundaresan2009b,Ma2020}.

Many models that explicitly address both RD and OD are of a single-compartmental and phenomenological nature~\cite{Hamlington2016,Mellenthin2019,Bates2020}. They allow for conclusions about general correlations, e.g., the categorization of lung injury by mechanical power dissipation and strain heterogeneity~\cite{Mellenthin2019}, or, that volu- and atelectrauma contribute to VILI in a combined manner and not independently~\cite{Hamlington2016}. Since neither local effects nor the patient-specific pathology are captured by these models, but these factors gain in importance when needing to adapt mechanical ventilation adequately~\cite{Kollisch-Singule2018,Gaver2020,Pelosi2021}, their benefit to clinical therapy management is limited. 

In addition, the few multi-compartment models of the lung mimicking RD and OD have shortcomings, e.g., due to not considering local pathology and neglecting time dependence of RD~\cite{Salmon1981,Sundaresan2009b,Amini2017a}. Especially the latter is known as a relevant phenomenon of RD~\cite{Bates2002,Albert2009}, and recent studies have demonstrated that timing has a great impact on the ventilation of injured lungs~\cite{Ramcharran2022,Nieman2020,Fardin2021,Gaver2020,Herrmann2021}.
In \textcolor{black}{a previous work}~\cite{Geitner2022}, we included the individual regional heterogeneity into a physics-based, anatomically accurate reduced dimensional lung model by 
incorporating time-dependent RD dynamics~\cite{Bates2002,Massa2008} on the conducting airway level and linking them to anatomical and pathological specifications. It was a first step toward including the potential for atelectrauma~(however without explicity addressing and investigating it) as well as volutrauma by locating and characterizing mechanical stress foci more accurately. Nevertheless, the research revealed the challenge in determining RD parameters uniquely for a patient, considering that physical conditions are not accessible locally and might differ regionally in the organ, and also from one patient to another. The model also neglected the probable presence of alveolar RD, which can result in an inaccurate regional volume capacity and, thus, deficient estimation of local straining on the one hand, and the cause of atelectrauma in the lung parenchyma on the other hand.

The current objective in this paper is to overcome these issues and advance the personalization of computational lung models by providing a tool \textcolor{black}{that can be} used to assess the VILI potential of certain ventilation profiles for a patient. We introduce a novel approach to modeling pressure- and time-dependent alveolar RD dynamics integrated into a viscoelastic component representing pulmonary tissue. One single (de-)recruitable viscoelastic component can represent both alveolar distension and RD. The idea being presented is motivated by previous studies~\cite{Hamlington2016,Mellenthin2019,Bates2020,Albert2009}, but it is based more on mechanical than phenomenological principles via the concept of modifying the stress-free reference volume of tissue in order to mimic RD, and it is meant to be applied as a multi-compartment model. This new model for RD is employed in the tissue elements (terminal units) of our comprehensive lung model. To enable patient-specific model calibration, we present a generic method that tailors the numerous model parameters (especially those related to RD) to ARDS patients based on ventilation characteristics and on their pathology extracted from medical image data. By applying this procedure to a specific patient, we examine the model's capacity to reproduce the clinical ventilation protocol performed at the bedside, and, thus, its predictive capability for this patient. Further, we investigate local straining and RD, the two mechanisms predominantly responsible for VILI. In the long run, such a digital twin can help to evaluate the injury potential of ventilation protocols for each patient specifically, and to eventually minimize VILI individually.

\section{Alveolar recruitment / derecruitment}
\label{sec:mat&methods}

\subsection{Physiological background}
\label{sec:characteristicsRD}

Lung tissue is a very delicate structure when viewing the small airways, alveolar ducts, and alveoli. The actual state of diseased tissue in vivo is very complex and remains difficult to resolve at the micro level. It can exhibit air-less and potentially recruitable lung units, air spaces flooded with inflammatory fluid, abnormal swelling of the alveolar wall, or a combination of these~\cite{Gattinoni2001}. 
Medical imaging enables identification of these scattered pathological regions, but their exact condition cannot be determined~\cite{Gattinoni2001}. 

Derecruitment due to lung edema may reduce the capacity of the tissue to distend because of fluid occupation or stiffening of alveolar walls, and can be regarded accordingly when modeling a patient's lung (e.g.,~\cite{Roth2017a}). In contrast to the relatively constant condition of edema presence, the phenomenon of cyclic intra-tidal opening and closing of lung structures has a very dynamic nature, which has been researched intensely. \textcolor{black}{Ghadiali and Huang}~\cite{Ghadiali2011} presented a comprehensive review on findings about RD and its impact on VILI. We will therefore only recapitulate herein the main characteristics of alveolar RD found in the literature and forming the basis of our modeling approach and underlying assumptions.

Alveolar ducts, and therefore alveolar tissue, change in size not only due to straining, but also because of (de-)folding and closure or opening~\cite{Knudsen2018a}. The collapse of injured distal airspaces happens heterogeneously~\cite{Knudsen2018a},
so not all at once on the acinar level. 
This (gradual) derecruitment due to closure of lung units reduces the amount of tissue available for distension during a breath cycle, thus resulting~---~precisely as in edema formation~---~in a decrease in lung compliance (meaning the change in lung volume per change in transpulmonary pressure)~\cite{Gattinoni2001,Gattinoni1987,Gattinoni2016}. Conversely, recruitment brings about an increase in lung compliance. 

To the extent evident, RD is primarily triggered by pressure, but exhibits a pronounced time-dependent behavior~\cite{Ramcharran2022,Fardin2021,Bates2002,Albert2009,Marini2008}. The pressure and time dependence each have individual characteristics: 

\begin{itemize}

    \item The prevailing pressure in the lung defines the eventual amount of recruited volume in a lung (in the subsequent paragraph denoted as $V_{\mathrm{final}}$)~\cite{Albert2009}. Further, the opening pressure of a lung unit is typically higher than the pressure at which a fully open and thus stable alveolar structure is supposed to close~\cite{Kollisch-Singule2020,Salmon1981,Hamlington2016,Massa2008,Frazer1985}. Viewing RD in an injured lung, the critical opening and closing pressures are not necessarily statistically distributed, but significantly scattered~\cite{Fardin2021}.

    \item The time dependence of RD has been observed in various experimental and clinical setups~\cite{Ramcharran2022,Fardin2021,Albert2009,Markstaller2001,Arnal2011,Pulletz2012}. The volume increase over time due to recruitment has often been described by a relaxation relationship following $V(t) = V_{\mathrm{final}} \left( 1-e^{-t/\tau} \right)$~\cite{Albert2009,Markstaller2001,Arnal2011,Pulletz2012}, with the final volume~$V_{\mathrm{final}}$ approximated over time during constant ventilation pressure and the time constant~$\tau$ specifying the speed of opening toward $V_{\mathrm{final}}$. \\\textcolor{black}{Previous studies}~\cite{Markstaller2001,Arnal2011,Pulletz2012}~employed this equation to specify the chronological progression of globally observed RD in ARDS lungs. Using the same relation in our model (see Section~\ref{sec:mat&methods:TUs}) provides a valuable indication for a reasonable choice of $\tau$ when tailoring the model parameters to a specific patient. 
    Further, \textcolor{black}{Albert~et~al.}~\cite{Albert2009}~showed in mice with saline lavaged lungs that the major alveolar recruitment happens after the first two seconds, regardless of the prevailing pressure level, i.e., a difference in the recruitment pressure level did not influence the time constant~$\tau$. 
    \\\\\textit{Remark: In a diseased lung, there are supposedly two different types of time dependencies, i.e.,~viscous effects that originate from normal tissue expansion and those that arise due to RD~\cite{Mellenthin2019,Pulletz2012}. Our model accounts for both types of time dependencies.}
    
\end{itemize}

\subsection{Modeling of alveolar recruitment/derecruitment}
\label{sec:mat&methods:TUs}

We use a well-established model element for acinar tissue~\cite{Roth2017a,Ismail2013,Roth2017} and enhance it with a novel, mechanically motivated model for alveolar RD based on the RD characteristics presented, which we will describe in greater detail hereinafter. Given that this model is eventually used at the outlets of conducting reduced dimensional airway elements (see underlying model in Section~\ref{sec:mat&methods:airways} in the Appendix) and mimics the subsequent acinar region, we will hereinafter refer to this tissue element as the \textit{terminal unit}. 

\begin{figure}[ht]
    \centering
    \includegraphics[page=1,trim=230 240 500 60,clip,width=0.25\textwidth]{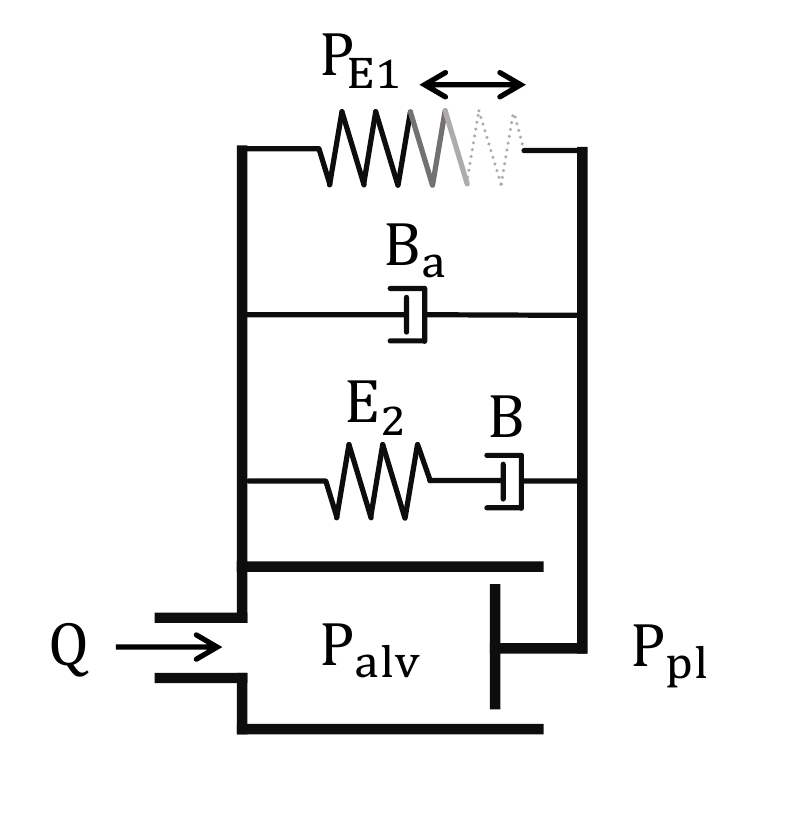}
    \caption{Schematic representation of the four-element Maxwell model underlying the terminal units. The non-linear spring~$P_\mathrm{E1}$ has a variable length which represents the concept of RD by variation of the stress-free reference volume achieved in the present research. See the text herein for a specific definition of all variables and parameters.}
    \label{fig:maxwellmodel}
\end{figure}
The viscoelastic behavior of lung parenchyma can be described in a reduced-dimensional manner by the arrangement of springs and dashpots (generalized Maxwell model~\cite{Bates}). In this study, we assume every terminal unit to consist of $N_\mathrm{ad}$~alveolar ducts \textcolor{black}{(Figure~\ref{fig:maxwellmodel})}, each represented by a Kelvin-Voigt body (spring and dashpot connected in parallel) in parallel with a Maxwell body (spring and dashpot connected in series)\textcolor{black}{~\cite{Roth2017a,Geitner2022,Ismail2013,Roth2017}}.
A terminal unit then follows
%
%
\begin{equation}
    \begin{split}
        N_\mathrm{ad} P + N_\mathrm{ad} \dfrac{B}{E_\mathrm{2}} \left( \dfrac{dP}{dt} \right) = \left( \dfrac{B B_\mathrm{a}}{E_\mathrm{2}} \right) \left( \dfrac{dQ}{dt} \right) + \left( B + B_\mathrm{a} \right) Q + N_\mathrm{ad} \dfrac{B}{E_\mathrm{2}} \left( \dfrac{dP_\mathrm{E1}}{dt} \right) + N_\mathrm{ad} P_\mathrm{E1},
    \end{split}
    \label{eq:maxwell_flow}
\end{equation}
where $Q$ is the gas flow into the terminal unit, $P$ is the pressure difference between the alveolar pressure~$P_{\mathrm{alv}}$ inside the terminal unit and the surrounding pressure~$P_{\mathrm{pl}}$ (see Section~\ref{sec:mat&methods:pressBC} for more details), $B$ and $B_\mathrm{a}$ are linear dashpots modeling time-dependent effects of the viscoelastic tissue distension, and $P_\mathrm{E1}$ and $E_\mathrm{2}$ are the non-linear and linear springs, respectively.
Especially in the context of OD, mimicking the non-linear behavior of lung tissue is a crucial modeling aspect. This effect is achieved by the non-linear spring~$P_\mathrm{E1}$ which represents the static pressure-volume relationship of a terminal unit due to its arrangement in the generalized Maxwell model. 
The non-linear pressure-volume behavior of $P_\mathrm{E1}$ is derived from a purely volumetric deformation of an Ogden-type material~\cite{Roth2017a,Geitner2022,Ogden1972} yielding
\begin{equation}
    \begin{split}
        P_\mathrm{E1} 
        = \frac{\kappa}{\beta} \cdot \frac{V_{\mathrm{0}}}{V} \left(1 -  \left( \frac{V_{\mathbf{0}}}{V} \right)^{\beta} \right)
    \end{split}
    \label{eq:Ogden}
\end{equation}
where $V$ denotes the current gas volume of a terminal unit and $V_{\mathrm{0}}$ the reference value of $V$ in the stress-free state. $\kappa$ and $\beta$ are slope- and curvature-shaping parameters, respectively.
\begin{figure*}[ht]
    \begin{center}
    \includegraphics[trim=0 30 0 100,clip,width=1.0\textwidth]{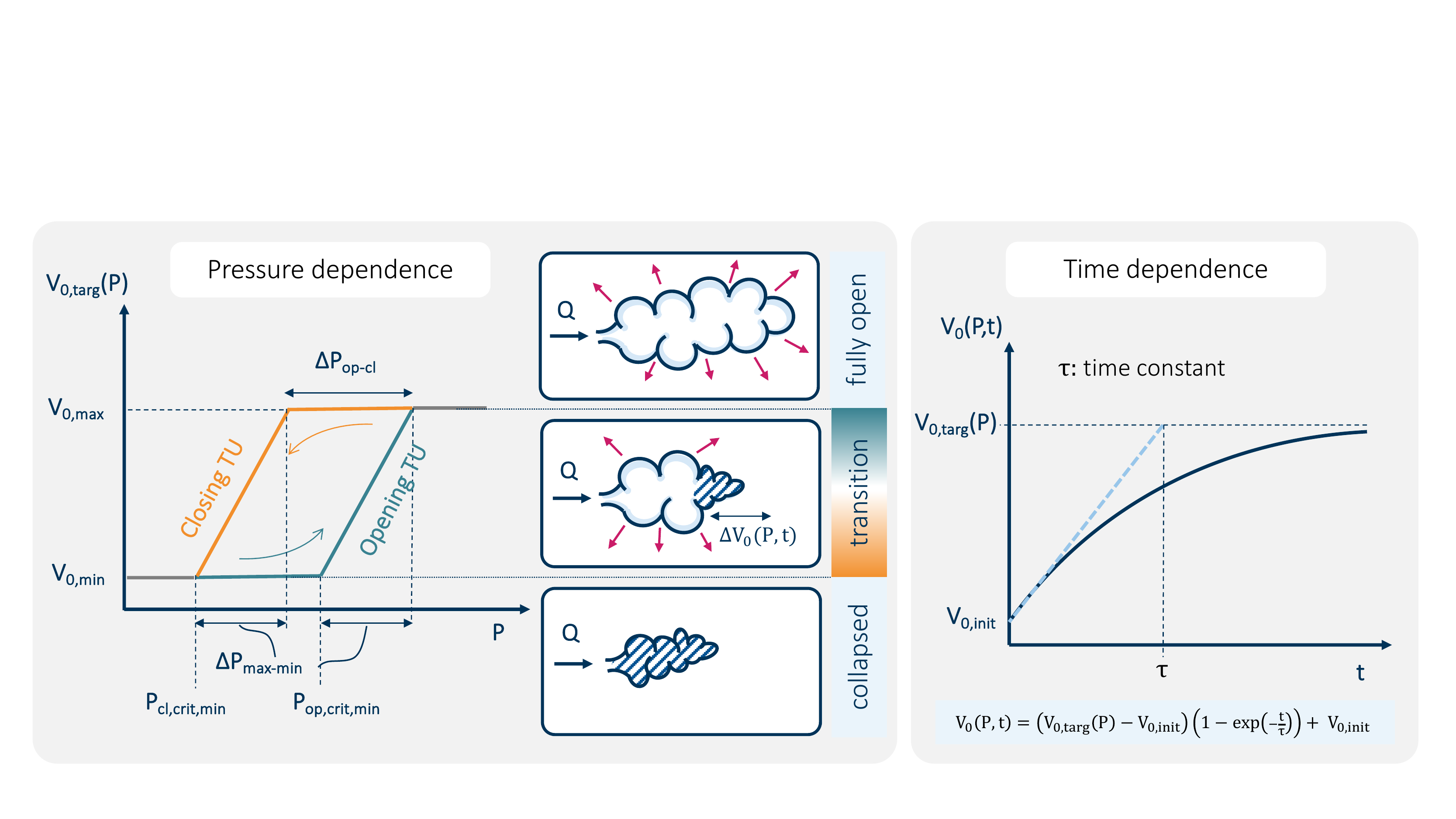}
    \caption{Pressure and time dependent collapse dynamics of the alveolar RD model presented. \textit{Pressure dependence:} The targeted reference volume~$V_{\mathrm{0,targ}} \left( P \right)$ of a terminal unit lies between the minimal and maximal reference volumes~$V_{\mathrm{0,min}}$ and $V_{\mathrm{0,max}}$, respectively, depending on the pressure difference~$P$ experienced by the terminal unit, and the active path, i.e., the opening~(blue) or closing path~(orange) specified by the minimal critical pressures for opening and closing, $P_{\mathrm{cl,crit,min}}$ and $P_{\mathrm{op,crit,min}}$, and the pressure offsets $\Delta P_{\mathrm{max-min}}$ and $\Delta P_{\mathrm{op-cl}}$. \textit{Time dependence:} Starting at an initial reference volume~$V_{\mathrm{0,init}}$, $V_{\mathrm{0,targ}} \left( P \right)$ is approximated over time~$t$ based on the time constant~$\tau$, which finally results in the current reference volume $V_{\mathrm{0}}\left( P,t \right)$. $\tau$ determines the time delay of the change in $V_{\mathrm{0}}\left( P,t \right)$, and can be visualized by the intersection of the slope of the reference volume at $t = 0s$ (blue dashed line) and $V_{\mathrm{0,targ}} \left( P \right)$. See the text and nomenclature for further explanation and a specific definition of all variables and parameters.}
    \label{fig:collapse_dynamics}
    \end{center}
\end{figure*}

In the case of (partially and progressively) collapsing alveolar structures or the infiltration of inflammatory liquid and edema, we assume that the alveolar volume available for expansion reduces, as outlined above. Similarly, the recruitment of air spaces increases the available volume. These changes in volume eventually decrease or increase the overall compliance of a terminal unit, respectively.
\\Considering both this effect and the time and pressure dependency of RD, we included RD dynamics in the model of a terminal unit presented above by making $V_{\mathrm{0}}$ in Eq.~\eqref{eq:Ogden} dependent on pressure and time~(Figure~\ref{fig:collapse_dynamics}), which is referred to as the current reference volume~$V_{\mathrm{0}}(P,t)$. 

\subsubsection{Pressure dependence}
The $P$ experienced by a terminal unit regulates $V_{\mathrm{0}}$ as follows:
Below a minimal critical pressure~$P_{\mathrm{crit,min}}$, the reference gas volume tends to fully collapse and reach its minimal volume~$V_{\mathrm{0,min}}$~(Figure~\ref{fig:collapse_dynamics}, left). Above $P_{\mathrm{crit,min}}$, the target value of the reference gas volume linearly depends on $P$ until it reaches the maximal critical pressure $P_{\mathrm{crit,max}} = P_{\mathrm{crit,min}} + \Delta P_{\mathrm{max-min}}$. Above $P_{\mathrm{crit,max}}$, $V_{\mathrm{0}}$ approaches the maximal reference gas volume $V_{\mathrm{0,max}}$, indicating a fully open state of all alveoli. The relation between $V_{\mathrm{0}}$ and its driving pressure can be formulated according to
\begin{equation}
    V_{\mathrm{0,targ}} \left( P \right) =
    \begin{cases}
      V_{\mathrm{0,min}} & \text{if $P < P_{\mathrm{crit,min}}$} \\
      m \cdot \left( P - P_{\mathrm{crit,min}} \right) + V_{\mathrm{0,min}} & \text{if $P_{\mathrm{crit,min}} \leq P \leq P_{\mathrm{crit,min}} + \Delta P_{\mathrm{max-min}}$} \\
      V_{\mathrm{0,max}} & \text{if $P > P_{\mathrm{crit,min}} + \Delta P_{\mathrm{max-min}}$}
    \end{cases}    
    \label{eq:ref_vol}
\end{equation}
with 
\begin{equation}
    m = \frac{V_{\mathrm{0,max}} - V_{\mathrm{0,min}}}{\Delta P_{\mathrm{max-min}}},
    \label{eq:ref_vol_slope}
\end{equation}
where the pressure dependent reference volume calculated from Eq.~\eqref{eq:ref_vol} is denoted as target reference volume $V_{\mathrm{0,targ}} \left( P \right)$, to which $V_{\mathrm{0}}(P,t)$ tends in a time dependent manner. The time dependence of $V_{\mathrm{0}}(P,t)$ is described in the following paragraph.

This approach makes it possible to model not only an open and closed state \textcolor{black}{in a binary manner~\cite{Geitner2022}} when using a RD model in the conducting airways, but also the gradual RD and partial collapse of the heterogeneously (de-)recruiting alveolar structure represented by one terminal unit. In this state, the open tissue proportion can still expand and potentially also experience OD. The current reference volume $V_{\mathrm{0}}(P,t)$ also provides an indication of the opening degree of a terminal unit.
To supplement the RD model by also considering that critical opening pressures usually exceed the closing pressures of a stable or recruited alveolar structure (see Section~\ref{sec:characteristicsRD}), we introduce two different paths for the recruitment and derecruitment process of a terminal unit, which differ by a shift $\Delta P_{\mathrm{op-cl}}$ of the corresponding critical pressures (see opening and closing path in Figure~\ref{fig:collapse_dynamics},~left).

\textcolor{black}{Due to the lack of knowledge on the shape of the pressure-volume relationship for opening of a conglomerate of alveoli, the assumption of a linear relationship between reference gas volume and applied pressure~$P$ is a first starting point. To the best of our knowledge, the literature on the underlying characteristics is very sparse. Thus, this approach can undoubtedly be refined in future work as more insights become available from experiments or resolved simulations.}

\subsubsection{Time dependence}
As introduced in Section~\ref{sec:characteristicsRD}, we used a time dependence relationship previously applied in a variety of contexts in order to describe the temporal alveolar (de-)recruitment in injured lungs:
$V_{\mathrm{0}} \left( P,t \right)$ yields $V_{\mathrm{0,targ}} \left( P \right)$ over time according to
\begin{equation}
    V_{\mathrm{0}}\left( P,t \right) = \left( V_{\mathrm{0,targ}}\left( P \right) - V_{\mathrm{0,init}} \right) \left( 1 - e^{-t/\tau} \right) + V_{\mathrm{0,init}},
    \label{eq:ref_vol_time}
\end{equation}
where $V_{\mathrm{0,init}}$ is an initial reference gas volume and $\tau$ is a time constant specifying the delay in RD of the terminal unit
. Since $V_{\mathrm{0}} \left( P,t \right)$ tends to $V_{\mathrm{0,targ}} \left( P \right)$ without ever reaching it mathematically, we introduce the tolerance factor~$\epsilon_\mathrm{V_\mathrm{0}}$ to ensure that a terminal unit can be opened or closed entirely. If a previously closed terminal unit (i.e., with a $V_{\mathrm{0,targ}} \left( P \right)$ moving on the opening path, see Figure~\ref{fig:collapse_dynamics}, left) exceeds $\left( 1 - \epsilon_\mathrm{V_\mathrm{0}} \right) V_{\mathrm{0,max}}$, it is assumed to fully recruit, and the opposite is assumed when falling below $ \left( 1 + \epsilon_\mathrm{V_\mathrm{0}} \right) V_{\mathrm{0,min}}$. Such a change of state eventually triggers the switch between the opening and closing paths and, therefore, between the relevant critical pressures.\\\\
Due to their clustered structure, we assume that the $N_\mathrm{ad}$ alveolar ducts of a single terminal unit in Eq.~\eqref{eq:maxwell_flow} have similar RD parameters used in Eqs.~\eqref{eq:ref_vol}~-~\eqref{eq:ref_vol_time} modeled by a linear relation between $P$ and $V_\mathrm{0,targ}$ for the entire terminal unit. This approach is based on the assumption that the subregion of the respiratory zone modeled by a terminal unit is governed by a specific, relatively uniform (patho-) physiological condition. One method used to determine the parameters for a single terminal unit based on patient imaging and ventilation data is described in Section~\ref{sec:mat&methods:parametrization}, resulting in individual parameter sets for each element in the framework of a multi-compartment patient-specific lung model.

As described in Section~\ref{sec:characteristicsRD}, alveolar derecruitment might also occur due to infiltration or edema in the alveolar tissue. This occupation of air spaces eventually decreases the distensible volume. In this study, we account for the accumulated fluid in the RD model by a constant amount of volume by which $V_{\mathrm{0,max}}$ is reduced (Section~\ref{sec:mat&methods:parametrization}).

We implemented the presented model by numerically discretizing Eqs.~\eqref{eq:maxwell_flow}~-~\eqref{eq:ref_vol_time} with the first-order Euler scheme. Similar to \textcolor{black}{Bates and Irvin}~\cite{Bates2002}, we performed several tests in an idealized setup to confirm that the novel RD model for terminal units can reproduce typical RD characteristics, and evaluated the influence of individual model parameters of Eqs.~\eqref{eq:ref_vol}~-~\eqref{eq:ref_vol_time}. However, in order to not overload this paper we refrain from showing the underlying tests.


\section{Setting up a patient-specific model}
\label{sec:mat&methods:modelconcrete}

In the long run, one objective of the proposed alveolar RD model is that of being applicable to patient-specific comprehensive multi-compartment lung models with spatial resolution. In addition to the anatomically accurate geometry generation, this objective also requires the ability to realistically tailor all underlying parameters to a patient, especially the numerous RD parameters for each terminal unit. The parametrization should be feasible with a reasonable effort and include the ability to capture regional heterogeneity as observable in medical image data. 
For this purpose, we developed a generic procedure that considers a patient's pathology, as extracted from medical imaging, and the mechanical behavior of the patient's respiratory system obtained from a few clinical ventilation measurements.
In the following, we describe the full approach to generating and calibrating an anatomically accurate, comprehensive multi-compartment lung model for a specific patient incorporating the presented model for alveolar RD.

\subsection{Clinical data}
\label{sec:mat&methods:data}

\textbf{Image data}\quad Generating the model of a patient-specific lung requires a three-dimensional thoracic CT~scan of the ventilated patient recorded at a known level of~---~usually positive~---~end-expiratory pressure~(PEEP) along normal ventilation. Such a CT~scan is usually part of the standard protocol for ARDS patients when admitted to the ICU. Based on this scan, we extract the geometry and individual pathology of the lung indicated by gray values of the image voxels. \\
\textbf{Ventilation data}\quad To calibrate the model parameters, we use bedside ventilation measurements, including the pressure at the airway opening, the tracheal airflow entering the lung, and the esophageal pressure as a surrogate for the pleural pressure. We specifically also use measurement of a normal breath cycle of the ventilation mode during which the CT scan has been recorded, and a quasi-static inflation maneuver to obtain information about the mechanical properties of the patient's respiratory system. Transpulmonary pressures 
used in the following are computed as the difference between the pressure at the airway opening and the pleural pressure.

\subsection{Geometry generation}
\label{sec:mat&methods:geo}

The conducting airway tree of a lung model starts from the centerline of the tracheal tree which reaches to the lobar bronchi leading into the individual lung lobes. The geometry and dimension of all these components are extracted from the CT scan. Due to the limited resolution of the medical images, we apply a space-filling tree growing algorithm for the regions \textcolor{black}{beyond the lobar bronchi~\cite{HowatsonTawhai2000}}, and extended and used in several of our previous studies~\cite{Roth2017a,Ismail2013,Roth2017}. A single airway segment of the conducting airway tree is modelled as described in Section~\ref{sec:mat&methods:airways} in the Appendix. Every terminal airway of the tree supplies gas to a terminal unit attached to it (see underlying model in Section~\ref{sec:mat&methods:TUs}). 

\begin{figure*}[ht]
    \begin{center}
    \includegraphics[trim=0 90 80 90,clip,width=1.0\textwidth]{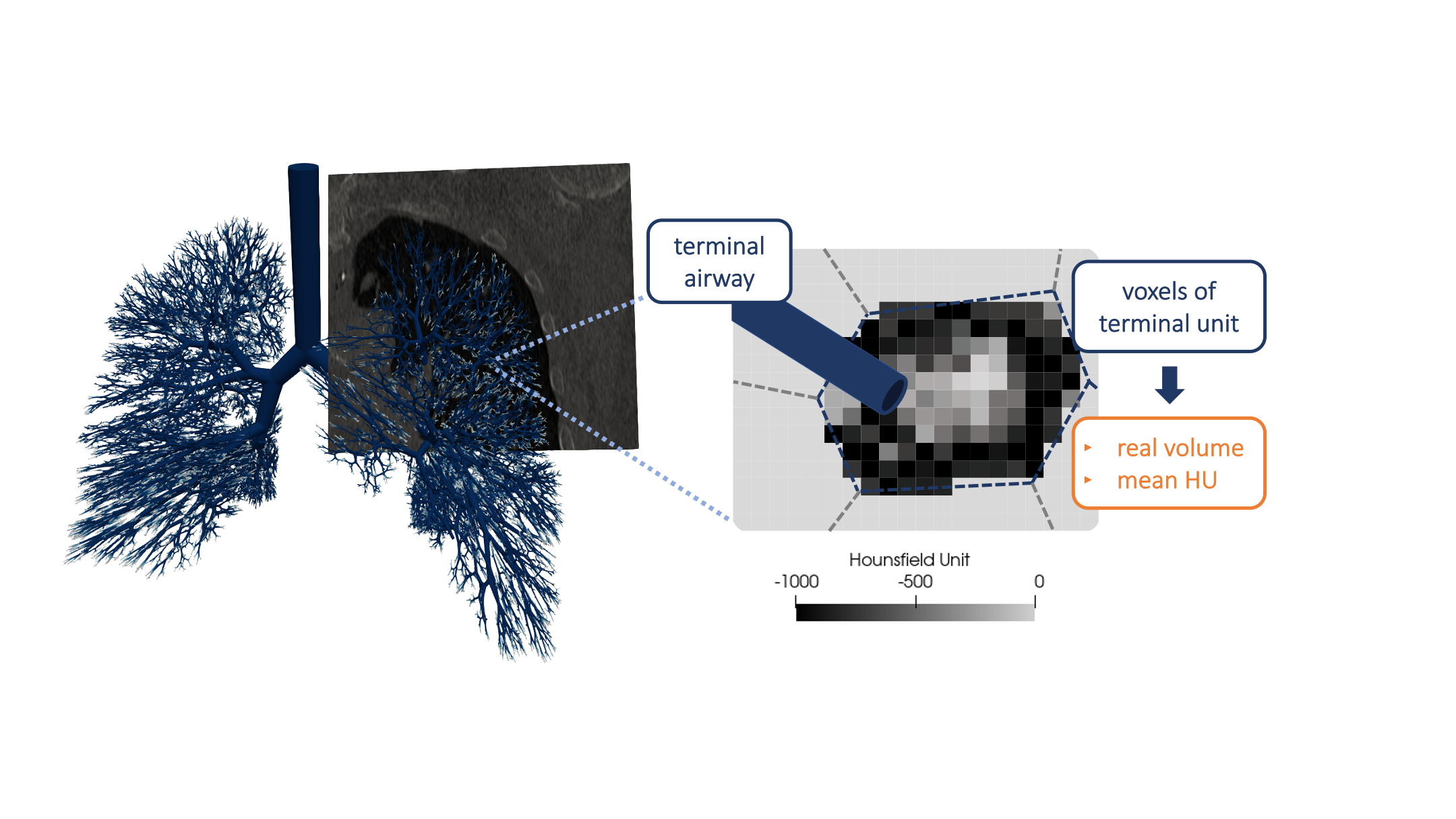}
    \caption{Conducting airway tree of a computational lung model resulting from the space-filling tree growing algorithm, illustrated using a slice from the underlying CT scan~(left); the voxel cloud remaining at a terminal airway after continuous splitting during the tree growing algorithm~(right) defines the water and gas volume and the mean HU of a terminal unit.}
    \label{fig:tree}
    \end{center}
\end{figure*}

To assign a realistic volume and Hounsfield~Unit~(HU) to each terminal unit, we slightly extended the tree growing procedure: The cloud of voxels constituting the lung region in the CT scan is continuously split into subgroups with proceeding branching of the airway tree. All voxels which remain at a terminal airway are assigned to the adherent terminal unit (Figure~\ref{fig:tree}). We can in this way attribute a total volume and a mean HU~---~calculated from the voxels~---~to each terminal unit. This information enables characterization of the composition of the total volume of gas ($V_{\mathrm{PEEP}}$) and water ($V_{\mathrm{water}}$) following HU~=~-1000 equals air and HU~=~0 equals water ($\sim$~tissue and edema liquid)~\cite{Gattinoni2001}. These quantities are used to parametrize the novel RD model in Section~\ref{sec:mat&methods:parametrization}. 

\subsection{Individualized pressure boundary conditions}
\label{sec:mat&methods:pressBC}

\begin{figure}[ht]
    \centering
    \includegraphics[page=3,trim=300 180 250 50,clip,width=0.50\textwidth]{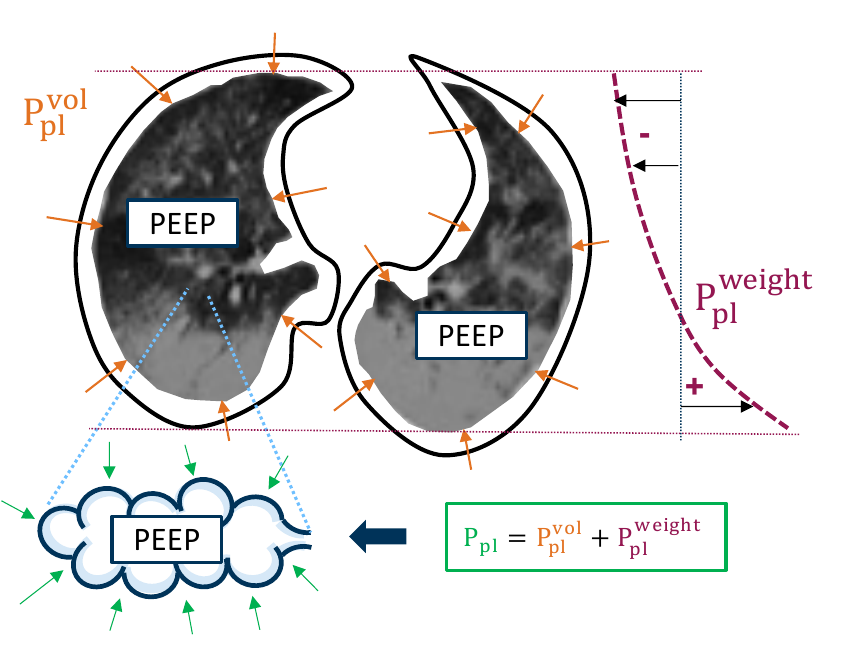}
    \caption{Representation of the height-dependent pressure state of terminal units at PEEP, being $P_{\mathrm{tp,PEEP}} = \mathrm{PEEP} - P_{\mathrm{pl}}$; the external pressure $P_{\mathrm{pl}}$ consists of a volume-dependent component $P^{\mathrm{vol}}_{\mathrm{pl}}$ and a height-dependent superimposed pressure $P^{\mathrm{g}}_{\mathrm{pl}}$ determined from the CT scan~\cite{Pelosi1994}.}
    \label{fig:collapse_dynamics_params_pressurestate}
\end{figure}

\textcolor{black}{Building on our previous lung model~\cite{Geitner2022}}, but with slight detail modifications, the external pressure~$P_{\mathrm{pl}}$ acting on the terminal units in simulations consists of two components: the variable, volume-dependent pressure~$P_{\mathrm{pl}}^\mathrm{vol}$ and the static contribution~$P_{\mathrm{pl}}^\mathrm{weight}$ introducing the weight of the lung regions above the units in question as an additional super-imposed pressure, and thus depending on the height~(Figure~\ref{fig:collapse_dynamics_params_pressurestate})~\cite{Pelosi1994,Crotti2001}. 
Hence, being the sum of these two constituents, the pleural pressure reads $P_{\mathrm{pl}} = P_{\mathrm{pl}}^\mathrm{vol} + P_{\mathrm{pl}}^\mathrm{weight}$. 

Basically, $P_{\mathrm{pl}}^\mathrm{vol}$ is the passive pressure exerted by the sedated chest wall of a patient against the expanding lung. In contrast to \textcolor{black}{our previous work}~\cite{Geitner2022}, we use a non-linear progression for the relationship between lung volume and pleural pressure reading
\begin{align}
P_{\mathrm{pl}}^{\mathrm{vol}} = a_\mathrm{v} + b_\mathrm{v} \cdot V_{\mathrm{frac}} + c_\mathrm{v} \cdot e^{d_\mathrm{v} \cdot V_{\mathrm{frac}}} ,
\label{eq:pleural_pressure_volume_dep} 
\end{align}
with 
\begin{align}
V_{\mathrm{frac}} = \frac{(V_{\mathrm{tot}} - V_{\mathrm{tot,PEEP}})}{(V_{\mathrm{tot,max}} - V_{\mathrm{tot,PEEP}})}, \:
\end{align}
where $a_\mathrm{v}$, $b_\mathrm{v}$, $c_\mathrm{v}$ and $d_\mathrm{v}$ are determined for the individual patient. $V_{\mathrm{frac}}$ is the volume share of the volume increase in all terminal units, $V_{\mathrm{tot}} - V_{\mathrm{tot,PEEP}}$, from the overall air volume in the lung at PEEP, $V_{\mathrm{tot,PEEP}}$, calculated from the CT scan, and the increase of air volume at end-inspiration of the measured quasi-static pressure-volume curve of the patient, $V_{\mathrm{tot,max}} - V_{\mathrm{tot,PEEP}}$.

In diseased lungs, the pressure gradient occurring across the lung due to its own weight is often more pronounced due to additional weight from pathological fluid accumulations in the organ~\cite{Pelosi1994}. For determination of the patient-specific $P_{\mathrm{pl}}^{\mathrm{weight}}$, which is in the patient's supine position a function of the ventral-to-dorsal height of the lung, $h$, we largely followed the approach used by Pelosi~et~al.~\cite{Pelosi1994}. Instead of dividing the lung only into ten vertical intervals, however, we discretize the lung in voxel slices to calculate the respective superimposed pressure in each layer. For the sake of precision, we fit the resulting discrete pressure values to a quartic relationship in place of a quadratic curve~\cite{Pelosi1994}, yielding
\begin{align}
P_{\mathrm{pl}}^{\mathrm{weight}} = a_{\mathrm{w}} + b_{\mathrm{w}} \left( h - h_{\mathrm{balloon}} \right) + c_{\mathrm{w}} \left( h^{2} - h_{\mathrm{balloon}}^{2} \right) + d_{\mathrm{w}} \left( h^{3} - h_{\mathrm{balloon}}^{3} \right) + e_{\mathrm{w}} \left( h^{4} - h_{\mathrm{balloon}}^{4} \right),
\label{eq:pressure_gradient_superimposed} 
\end{align}
where $a_{\mathrm{w}}$, $b_{\mathrm{w}}$, $c_{\mathrm{w}}$, $d_{\mathrm{w}}$ and $e_{\mathrm{w}}$ are fitting parameters. To enforce $P_{\mathrm{pl}}^{\mathrm{weight}} = 0$ at the reference point of measurement of $P_{\mathrm{pl}}$, which was made with the esophageal balloon in the esophagus (see positive and negative range of $P_{\mathrm{pl}}^{\mathrm{weight}}$ in Figure~\ref{fig:collapse_dynamics_params_pressurestate}), we amended Eq.~\eqref{eq:pressure_gradient_superimposed} with the height of the measurement spot, $h_{\mathrm{balloon}}$, determined from the CT scan~\cite{Yoshida2018}.

\subsection{Image- and ventilation-based parametrization of terminal units}
\label{sec:mat&methods:parametrization}

\textcolor{black}{In order to apply the presented model of alveolar RD to a real ARDS~lung and individualize it for a patient, we propose a sophisticated approach to calibrate the numerous model parameters. It bases on a deep understanding of the model and lung mechanics, and uses image data and deliberately selected ventilation maneuvers to extract certain patient-specific information. 
In this way, we manage to handle the underlying complexity when applying the model to a real case and establish an innovative way to determine the model parameters for the respiratory mechanics and lung pathology of a specific patient. 
For the sake of completeness, before going into more detail about the calibration method, we would like to point out that there are less refined and not explicitly designed for our purpose, but still effective procedures for model parametrization, e.g., Bayesian inverse analysis methods, which our group is currently developing also in the context of biomechanical problems~\cite{Hervas-Raluy2023,Willmann2022a,Nitzler2022}. These methods might be of particular interest when using data from further or different measurement sources that are not used in the calibration method described below.}

In the following, we propose a two layered procedure that parametrizes the terminal units: 
On the one hand, we determine the RD parameters in Eqs.~\eqref{eq:ref_vol}~--~\eqref{eq:ref_vol_time} for each terminal unit by means of an algorithm (Section~\ref{sec:mat&methods:parametrization:algo}) based on the mean gray value and the gas volume of the terminal unit assigned in the geometry generation, and based on its approximated height-dependent pressure state at end-inspiration and end-expiration during a measured normal ventilation breath cycle. The gray values herein indicate the pathological state of the terminal unit and thus enable adequate selection of its RD parameters to reproduce the regional heterogeneity of the lung.
On the other hand, the aforementioned algorithm requires a set of input parameters (see a list of the input parameters in Table~\ref{tab:input_params}), which are optimized to match the overall pressure-volume behavior of all terminal units to the measurements of a normal breath cycle and a quasi-static inflation maneuver (Section~\ref{sec:mat&methods:parametrization:opti}).

\paragraph{Assumptions}

To simplify the algorithm for RD parametrization and the optimization of its necessary input parameters, both of which are described in greater detail hereinafter, we for now base it on the following assumptions below. It is important to note that these assumptions are made only during model parametrization and do not hold for the whole computational model.

\begin{itemize}
    \item We disregard the effects of the upstream airway tree, e.g.,~the pressure drop across the conducting airways due to their flow resistance. We thus assume that the pressure at the airway opening directly acts upon the terminal units, especially at quasi-static points during ventilation like end-inspiration, end-expiration and during quasi-static inflation, where this assumption approximately holds true.
    \item We only consider the elastic component of each terminal unit, i.e.,~the non-linear spring~$P_\mathrm{E1}$ following the relation in Eq.~\eqref{eq:Ogden} that incorporates RD dynamics, but disregard any effects of the spring~$E_\mathrm{2}$ and the dashpots~$B$ and $B_\mathrm{a}$ in Eq.~\eqref{eq:maxwell_flow}.
    \item The pressure state within the lung during recording of the CT scan, $P_{\mathrm{tp,PEEP}}$, is $\mathrm{PEEP} - P_{\mathrm{pl}}$, with the latter depending from $V_{\mathrm{tot,PEEP}}$ and $h$ as depicted in Figure~\ref{fig:collapse_dynamics_params_pressurestate} and described in Section~\ref{sec:mat&methods:pressBC}. In other words, a specific and identifiable height-dependent $P_{\mathrm{tp,PEEP}}$ is assumed for each terminal unit at PEEP in the moment of capturing the CT scan which is relevant when determining the critical RD pressures~\cite{Crotti2001,Scaramuzzo2020}.
    \item We distinguish among three types of terminal units, depending on the HU assigned, as is similar to the assumed pathology condition in an ARDS~lung~\cite{Nieman2020,Gattinoni2001,Gattinoni1987}: normally aerated (including hyper-ventilated) terminal units with $\mathrm{HU} \leq -500$, poorly ventilated terminal units with $ -500 < \mathrm{HU} < -100 $, and non-aerated terminal units with $\mathrm{HU} \geq -100 $.
    \item We assume $\tau = \tau_\mathrm{insp} = \tau_\mathrm{exp}$ of a terminal unit for the sake of simplicity in contrast to \textcolor{black}{the model proposed by Bates~and~Irvin~\cite{Bates2002} in which} airways tended to close faster than they opened. 
    \item For each collapsed and poorly ventilated terminal unit, we assume a predefined constant ratio~$k_\mathrm{edema}$ of edema volume and the thereby filled tissue volume. Based on the volumes $V_{\mathrm{water}}$ and $V_{\mathrm{PEEP}}$ extracted from the CT scan, we assume for $V_{\mathrm{tissue}}$ that it is comprised of tissue volume filled by edema fluid, and open tissue volume holding $V_{\mathrm{PEEP}}$.
\end{itemize}
\begin{figure}[ht]
    \centering
    \includegraphics[page=3,trim=300 150 300 100,clip,width=0.40\textwidth]{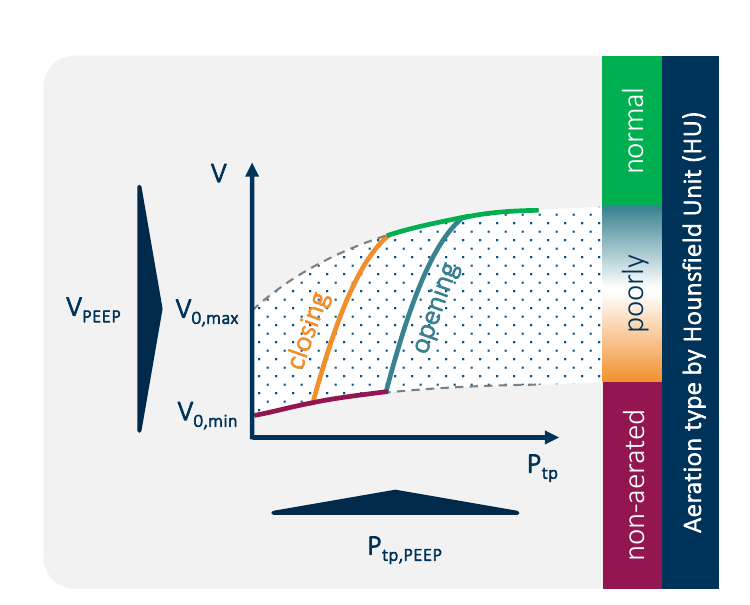}
    \caption{Determination of the location of a terminal unit on the pressure-volume curve and its RD parameters depending on the pathological condition of the terminal unit (indicated by HU) and its individual pressure and volume state at PEEP, i.e., the height-dependent $P_{\mathrm{tp,PEEP}}$ and the corresponding gas volume $V_{\mathrm{PEEP}}$ extracted from the CT scan.}
    \label{fig:collapse_dynamics_params_ac_categories}
\end{figure}

\subsubsection{Algorithm for parametrization}
\label{sec:mat&methods:parametrization:algo}

\paragraph{Determination of RD volumes}
In a first step, the algorithm determines $V_{\mathrm{0,min}}$ and $V_{\mathrm{0,max}}$ of each terminal unit depending on the HU~type and its location on the pressure-volume curve at PEEP ~(Figure~\ref{fig:collapse_dynamics_params_ac_categories}).

Normally aerated terminal units are assumed to be in a fully open state, i.e., their expansion from PEEP happens only due to distension and without any recruitment. Therefore, their current reference volume assumed from the CT scan at PEEP,~$V_{\mathrm{0,PEEP}}$, equals $V_{\mathrm{0,max}}$ (green curve in Figure~\ref{fig:collapse_dynamics_params_ac_categories}) and can be determined by solving Eq.~\eqref{eq:Ogden} for $P_{\mathrm{tp,PEEP}}$ and the known gas volume $V_{\mathrm{PEEP}}$. $\kappa$ and $\beta$ in Eq.~\eqref{eq:Ogden} are input parameters of the algorithm, i.e., presumed quantities that are fit in the outer optimization as described in Section~\ref{sec:mat&methods:parametrization:opti}.

The poorly and non-aerated terminal units are (partially) collapsed and assumed to be located somewhere between a fully open and a fully closed state (blue or orange curve in Figure~\ref{fig:collapse_dynamics_params_ac_categories}). Given this assumption, we can make no direct conclusions about $V_{\mathrm{0,max}}$ based on the gas volume in the CT scan (only about $V_{\mathrm{0,PEEP}}$ of the terminal units by Eq.~\eqref{eq:Ogden}). We thus use their $V_{\mathrm{tissue}}$ as an indication for $V_{\mathrm{0,max}}$ that can be reached by a terminal unit in fully open state. Taking the ratios $V_{\mathrm{0,max}}/V_{\mathrm{tissue}}$ of all normally aerated terminal units, we determine the mean and standard deviation of their normal distribution and chose $V_{\mathrm{0,max}}$ for the (partially) collapsed terminal units randomly and according to the probabilistic distribution.

To complete the RD volume parameters, we further assume for all terminal units $V_{\mathrm{0,min}} = k_{\mathrm{coll}} \cdot V_{\mathrm{0,max}}$ with $k_{\mathrm{coll}}$ being a further input parameter to the parametrization algorithm.

\paragraph{Determination of RD time constants}
Injured lungs exhibit a change in aeration during expiration that is faster in the dorsal than in the ventral lung when in a supine position~\cite{Herrmann2021}. Since this different temporal behavior is very subtle and might also be attributed to the difference in viscoelastic tissue straining due to gravitation, we do not consider pathology or height in the choice of the RD time constants. The time constants $\tau$ of all terminal units are randomly chosen from a distribution. The type and the parameters of the distribution are input parameters of the algorithm and specifically determined for a patient.

\paragraph{Determination of critical pressures}
In addition to the volumes, each terminal unit also requires the specification of the critical pressures $P_{\mathrm{cl,crit,min}}$, $P_{\mathrm{cl,crit,max}}$,$P_{\mathrm{op,crit,min}}$, and $P_{\mathrm{op,crit,max}}$. Given the constant relationship between the critical pressures defined by $\Delta P_{\mathrm{max-min}}$ and $\Delta P_{\mathrm{op-cl}}$, they are all fixed as soon as the value of one of them is known (Figure~\ref{fig:collapse_dynamics}). $\Delta P_{\mathrm{max-min}}$ and $\Delta P_{\mathrm{op-cl}}$ are both input parameters to the parametrization algorithm. 

\begin{figure*}[ht]
    \begin{center}
    \includegraphics[trim=0 120 90 70,clip,width=0.8\textwidth]{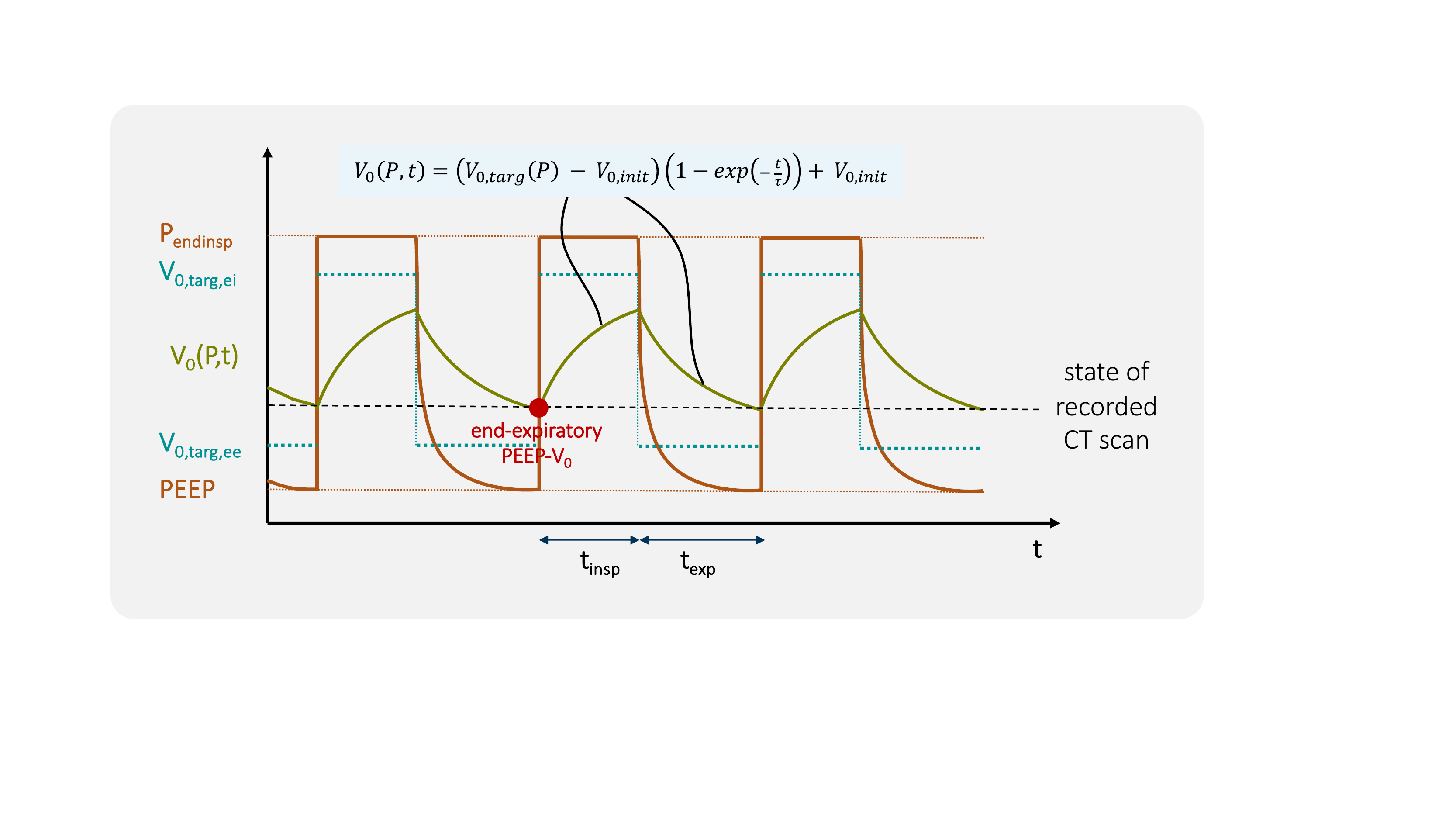}
    \caption{Assumed pressure- and time-dependent steady-state oscillation of the stress-free reference volume of a poorly ventilated terminal unit along normal ventilation cycles.}
    \label{fig:collapse_dynamics_params_timedep_steadystate}
    \end{center}
\end{figure*}

Before describing the procedure used to determine the critical pressures, we will offer a few basic thoughts:
Poorly ventilated terminal units are considered to be partially collapsed at the time of CT recording at PEEP, so they are assumed to take on an unstable state subjected to repetitive intra-tidal RD during normal ventilation. 
Therefore, the $V_{\mathrm{0,PEEP}}$ of a tissue element determined for $V_{\mathrm{PEEP}}$ and $P_{\mathrm{tp,PEEP}}$ by solving Eq.~\eqref{eq:Ogden}, lies somewhere between $V_{\mathrm{0,max}}$ and $V_{\mathrm{0,min}}$ and offers a clue about the critical pressures.
However, due to the time dependence of RD, $V_{\mathrm{0,PEEP}}$ is not the actual target volume~$V_{\mathrm{0,targ,ee}}$ pursued at $P_{\mathrm{tp,PEEP}}$ during normal ventilation, but the end-expiratory point of $V_{\mathrm{0}}(P,t)$ reached after the time of expiration $t_{\mathrm{exp}}$ during moving towards $V_{\mathrm{0,targ,ee}}$ (Figure~\ref{fig:collapse_dynamics_params_timedep_steadystate}). 
In a similar manner, during inspiration for the time $t_{\mathrm{insp}}$, $V_{\mathrm{0,targ,ei}}$ is approximated but not fully reached. 
The $V_{\mathrm{0}}(P,t)$ of a poorly ventilated terminal unit thus continuously moves in transient states during subsequent normal breath cycles and the state at end-expiration is gathered in the CT scan. The initial reference volumes $V_{\mathrm{0,init}}$ of a terminal unit at the beginning of inspiration or expiration remain unknown, and it is unclear whether they somehow achieve a steady state.

On the basis of the above consideration and open question, we derive the following mathematical relations in order to eventually determine the critical pressures of poorly ventilated terminal units.

Beginning at end-expiration, we take an initially unknown reference volume $V_{\mathrm{0,init}}$ of a terminal unit denoted as $V_{0,1}$ and assume a constant transpulmonary pressure $P_{\mathrm{tp,endinsp}} = P_{\mathrm{endinsp}} - P_{\mathrm{pl}}$, which includes the height-dependent gravitational load of a terminal unit, for the time of inspiration,~$t_{\mathrm{insp}}$. According to Eq.~\eqref{eq:ref_vol_time}, the new reference volume~$V_{0,2}$ at end-inspiration (and $V_{\mathrm{0,init}}$ for the following expiration) yields
\begin{align}
    V_{\mathrm{0,2}} = V_{\mathrm{0,1}} + \left(V_{\mathrm{0,targ,ei}} - V_{\mathrm{0,1}}\right) \left( 1 - e^{-t_{\mathrm{insp}}/\tau_{\mathrm{insp}}} \right).
\end{align}
Again, a constant $P_{\mathrm{tp,PEEP}}$ lasting for the time of expiration, $t_{\mathrm{exp}}$, results in a reference volume reading
\begin{align}
    V_{\mathrm{0,3}} = V_{\mathrm{0,2}} + \left(V_{\mathrm{0,targ,ee}} - V_{\mathrm{0,2}}\right) \left( 1 - e^{-t_{\mathrm{exp}}/\tau_{\mathrm{exp}}} \right).
\end{align}
Replacing the exponential expressions with the constants~$c_{\mathrm{I}} = e^{-t_{\mathrm{insp}}/\tau_{\mathrm{insp}}}$ and $c_{\mathrm{E}} = e^{-t_{\mathrm{exp}}/\tau_{\mathrm{exp}}}$, rearranging the equations, and continuing the stepwise analysis of consecutive inspiration and expiration phases leads to the general expression
\begin{equation}
    \begin{split}
    V_{\mathrm{0},n} = \: &c_{\mathrm{I}} \left(c_{\mathrm{I}} c_{\mathrm{E}} \right)^{\frac{n - 2}{2}} V_{\mathrm{0,1}} + 
    \left( \sum_{i=0}^{\frac{n - 2}{2}} \left( c_{\mathrm{I}} c_{\mathrm{E}} \right)^i \right) \left( 1 - c_{\mathrm{I}} \right) V_{\mathrm{0,targ,ei}}
    + \left( \sum_{i=0}^{\frac{n - 4}{2}} \left( c_{\mathrm{I}} c_{\mathrm{E}} \right)^i \right) c_{\mathrm{I}} \left( 1 - c_{\mathrm{E}} \right) V_{\mathrm{0,targ,ee}}
    \end{split}
    \label{eq:v0even}
\end{equation}
for even values of $n \geq 4$, with $V_{\mathrm{0},n}$ denoting the reference volumes at end-inspiration and in the following referred to as $V_{\mathrm{0,insp}}$, and
\begin{equation}
    \begin{split}
    V_{\mathrm{0},n} = \: & \left(c_{\mathrm{I}} c_{\mathrm{E}} \right)^{\frac{n - 1}{2}} V_{\mathrm{0,1}} + 
    \left( \sum_{i=0}^{\frac{n - 3}{2}} \left( c_{\mathrm{I}} c_{\mathrm{E}} \right)^i\right) c_{\mathrm{E}} \left( 1 - c_{\mathrm{I}} \right) V_{\mathrm{0,targ,ei}} \
    + \left( \sum_{i=0}^{\frac{n - 3}{2}} \left( c_{\mathrm{I}} c_{\mathrm{E}} \right)^i\right) \left( 1 - c_{\mathrm{E}} \right) V_{\mathrm{0,targ,ee}},
    \end{split}
    \label{eq:v0uneven}
\end{equation}
for uneven values of $n \geq 3$, with $V_{\mathrm{0},n}$ denoting the reference volumes at end-expiration and hereafter referred to as $V_{\mathrm{0,exp}}$.
%
Applying the general relation for geometric series $\sum_{i=0}^{\infty} x^i = \frac{1}{1 - x}$, which is always valid for $|x| < 1$ and, thus, true in our case ($t_{\mathrm{insp}}, t_{\mathrm{exp}} > 0$ and $ 0 < \tau_{\mathrm{insp}}, \tau_{\mathrm{exp}} < \infty$ resulting in $0 < c_{\mathrm{I}}, c_{\mathrm{E}} < 1$), we get
\begin{equation}
    \sum_{i=0}^{\infty} \left( c_{\mathrm{I}} c_{\mathrm{E}} \right)^i = \frac{1}{1 - c_{\mathrm{I}} c_{\mathrm{E}}}.
\end{equation}
Equations~(\ref{eq:v0even}) and (\ref{eq:v0uneven}) then converge for $n \rightarrow \infty$ to
\begin{equation}
V_{\mathrm{0,insp}} = \frac{1 - c_{\mathrm{I}} }{1 - c_{\mathrm{I}} c_{\mathrm{E}}}  V_{\mathrm{0,targ,ei}} + \frac{c_{\mathrm{I}} \left( 1 - c_{\mathrm{E}} \right)}{1 - c_{\mathrm{I}} c_{\mathrm{E}}} V_{\mathrm{0,targ,ee}},
\label{eq:v0insp}
\end{equation}
and
\begin{equation}
V_{\mathrm{0,exp}} = \frac{ c_{\mathrm{E}} \left( 1 - c_{\mathrm{I}} \right)}{1 - c_{\mathrm{I}} c_{\mathrm{E}}}  V_{\mathrm{0,targ,ei}} + \frac{ 1 - c_{\mathrm{E}}}{1 - c_{\mathrm{I}} c_{\mathrm{E}}} V_{\mathrm{0,targ,ee}},
\label{eq:v0exp}
\end{equation} 
respectively. Note that $V_{0,1}$ dropped out in these equations. Therefore, we reach a steady state oscillation between $V_{\mathrm{0,insp}}$ and $V_{\mathrm{0,exp}}$ that is independent from the initial reference volume assumed at the beginning. Obviously, the condition $n \rightarrow \infty$ is not met in reality. However, some iterative calculations demonstrated a good approximation of $V_{\mathrm{0,insp}}$ and $V_{\mathrm{0,exp}}$ after only a few breathing cycles for realistic mean values of $\tau$~\cite{Arnal2011,Pulletz2012}. A normal ventilation during CT recording lasting at least several minutes without changes in the ventilator setting should be a period of time long enough to approach this steady state acceptably.\\

$V_{\mathrm{0,exp}}$ in Eq.~\eqref{eq:v0exp} should now equal $V_{\mathrm{0,PEEP}}$ determined for all poorly aerated terminal units. For this purpose, we use a sampled range for reasonable critical pressures $P_{\mathrm{crit,max}}$ and $P_{\mathrm{crit,min}}$ and calculate $V_{\mathrm{0,targ,ei}}$ and $V_{\mathrm{0,targ,ee}}$ for each parameter set according to Eq.~\eqref{eq:ref_vol}, assuming constant $P_{\mathrm{tp,PEEP}}$ and $P_{\mathrm{tp,endinsp}}$ during expiration and inspiration, respectively. The parameter set where $V_{\mathrm{0,exp}}$ is the closest to $V_{\mathrm{0,PEEP}}$, derived from $V_{\mathrm{PEEP}}$ of the terminal unit, gathered in the CT scan, by solving Eq.~\eqref{eq:Ogden}, then holds as final $P_{\mathrm{crit,max}}$ and $P_{\mathrm{crit,min}}$.

Evaluating also $V_{\mathrm{0,insp}}$ helps to define the current path (opening or closing) an element moves on, which eventually sets $P_{\mathrm{crit,max}}$ and $P_{\mathrm{crit,min}}$ as $P_{\mathrm{op,crit,max}}$ and $P_{\mathrm{op,crit,min}}$, or $P_{\mathrm{cl,crit,max}}$ and $P_{\mathrm{cl,crit,min}}$, respectively.
If $V_{\mathrm{0}}$ reaches $V_{\mathrm{0,insp}} \geq \left( 1 - \epsilon_\mathrm{V_\mathrm{0}} \right) V_{\mathrm{0,max}}$ at end-inspiration but not $V_{\mathrm{0,exp}} \leq \left( 1 + \epsilon_\mathrm{V_\mathrm{0}} \right) V_{\mathrm{0,min}}$ at end-expiration, then the terminal unit is assigned to move on the closing path, and in all other cases on the opening path.\\

Regarding the fully open normally ventilated terminal units, we have no actual indications of the relevant critical pressures at hand, especially below which pressure they start to collapse. Thus, apart from ensuring $P_{\mathrm{cl,crit,max}} < P_{\mathrm{tp,PEEP}}$, their critical pressures are randomly chosen from a normal distribution with mean~$\mu_{\mathrm{cl}}$ and standard deviation~$\sigma_{\mathrm{cl}}$, two further input parameters of the global optimization loop. All normally ventilated terminal units are initially moving on the closing path.\\

The non-aerated terminal units are assumed to have a similar pathological condition to each other and, therefore, similar critical opening pressures. Thus, their critical opening pressures are randomly chosen from a normal distribution with the mean $\mu_{\mathrm{op}}$ and the standard deviation $\sigma_{\mathrm{op}}$~---~both again input parameters of the outer optimization loop. According to their aeration state at PEEP, we ensure that collapsed terminal units approximate a closed state at end-expiration by setting $P_{\mathrm{op,crit,min}} > P_{\mathrm{tp,PEEP}}$. Further, they are not supposed to open entirely during normal ventilation, guaranteed by $P_{\mathrm{op,crit,max}} > P_{\mathrm{tp,endinsp}}$. 

\subsubsection{Optimization of input parameters}
\label{sec:mat&methods:parametrization:opti}

As stated, the algorithm described above is based on a set of input parameters which are collected in Table~\ref{tab:input_params}. To adjust these quantities, the summed $P$-$V_{\mathrm{tot}}$ behavior of all terminal units is matched to the clinical measurements, i.e.,~to the $P$-$V_{\mathrm{tot}}$ points at end-inspiration and end-expiration of a normal breath cycle and to the quasi-static inflation maneuver. For the procedure, we again consider only the elastic component of the terminal units (Eq.~\eqref{eq:Ogden}) incorporating pressure- and time-dependent RD. The calibration process thereby also includes the intra-tidal time- and pressure-dependent RD of the terminal units which is continuously present in pathological tissue regions during ventilation~\cite{Albaiceta2004}, and should therefore be considered.

Specifically, we assume an ideal normal breath cycle at constant $P_{\mathrm{tp,PEEP}}$ and $P_{\mathrm{tp,endinsp}}$, specified by the ventilation measurements of the patient and acting on each terminal unit for the duration of $t_{\mathrm{exp}}$ and $t_{\mathrm{insp}}$, respectively. We calculate the gas volume $V$ of each terminal unit at end-expiration and end-inspiration which evolves due to the supposed pressure regime and the time, including RD effects. Finally, summing $V$ of all terminal units enables comparison of $V_{\mathrm{tot}}$ to the clinical measurement at the two quasi-static points.

This procedure is repeated for the quasi-static inflation maneuver, taking constant transpulmonary pressure courses finely discretized over time between the measured pressure level at the onset ($P_{\mathrm{tp,qs,start}}$), 
the peak inspiration ($P_{\mathrm{tp,qs,max}}$), and the end-expiration ($P_{\mathrm{tp,qs,end}} = P_{\mathrm{tp,PEEP}}$) of the quasi-static inflation maneuver. The time of inspiration $t_{\mathrm{insp,qs}}$ and time of expiration $t_{\mathrm{exp,qs}}$ of that maneuver is also taken from the clinical measurements. By calculating and summing the $V$ of all terminal units for every small time interval and the corresponding transpulmonary pressure, we produce an ideal pressure-volume curve including the time- and pressure dependent RD of the terminal units, which can be matched to the measured pressure-volume curve of the patient.

\section{Clinical example}

\subsection{Model setup}

\paragraph{Clinical data}

As part of generating an exemplary patient-specific model, we used chest CT images and measurements of a critically ill, endotracheally intubated male patient suffering from moderate ARDS. The patient was treated at the operative intensive care unit of the Department of Anesthesiology and Intensive Care Medicine at University Medical Center Schleswig-Holstein, Campus Kiel.
All data were provided in an anonymized format. Ethical approval was obtained from the ethics committee of the Medical Faculty in Kiel, and the underlying study was carried out in accordance with the Declaration of Helsinki. Written informed consent was obtained from the patient's legal representative.

The patient chest CT used was recorded at a PEEP of 8~mbar (PEEP8) with 512x512x339 pixels of 0.726562x0.726562x1~mm within 24~h before the start of the underlying study. An exemplary axial view of the lung excised from the CT scan is shown in Figure~\ref{fig:collapse_dynamics_params_pressurestate}. 

As part of the study protocol, various ventilation maneuvers were performed including the profiles required for the model calibration described in~\ref{sec:mat&methods:parametrization}. Remaining ventilation maneuvers that did not enter model calibration were used to validate the performance of the model in Section~\ref{sec:results}.

\paragraph{Model generation}

We used Mimics and 3-Matic (Materialise, Leuve, Belgium), to segment the patient-specific model geometry and extract dimensions of the trachea, the lobar bronchi, and the lung lobes from the medical images. The space-filling tree growing algorithm yielded the conducting airway tree shown in Figure~\ref{fig:tree}~(left), consisting of 58321 airway segments and having 29161 terminal units attached to the terminal airway segments. 

\paragraph{Model calibration}

\begin{figure}[ht]
\centering
\begin{minipage}{.45\textwidth}
    \centering
    \includegraphics[trim=0 0 0 0,clip,width=.9\textwidth]{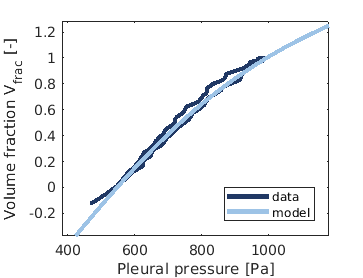}
\end{minipage}\qquad
\begin{minipage}{.45\textwidth}
    \centering
    \includegraphics[trim=0 0 0 0,clip,width=1.0\textwidth]{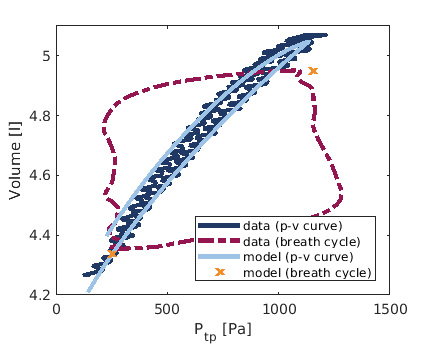}
\end{minipage}

\bigskip

\begin{minipage}[t]{.45\textwidth}
    \centering
    \caption{Relationship between the volume fraction and the pleural pressure used as boundary condition in the model (light blue) determined from the measured pleural pressure-volume curve (dark blue).}
    \label{fig:ppl_vol_fit}
\end{minipage}\qquad
\begin{minipage}[t]{.45\textwidth}
    \centering
    \caption{Idealized pressure-volume behavior (light blue solid) and quasi-static points (orange crosses) of all terminal units of the lung model resulting for the finally chosen input parameters of RD parametrization, compared to clinical measurements of the quasi-static inflation maneuver (dark blue solid) and a normal breath cycle (red dashed) .}
    \label{fig:fit_lowflow}
\end{minipage}

\end{figure}

We determined the parameters in the two components of $P_{\mathrm{pl}}$ for the patient in MATLAB by non-linear regression. Fitting Eq.~\eqref{eq:pleural_pressure_volume_dep} to the patient's pressure-volume data resulted in the curve of $P_{\mathrm{pl}}^{\mathrm{vol}}$ depicted in Figure~\ref{fig:ppl_vol_fit}. 
The fitted course of $P_{\mathrm{pl}}^{\mathrm{weight}}$ (Eq.~\eqref{eq:pressure_gradient_superimposed}) is qualitatively shown in Figure~\ref{fig:collapse_dynamics_params_pressurestate}. The chosen values for the parameters in $P_{\mathrm{pl}}^{\mathrm{vol}}$ and $P_{\mathrm{pl}}^{\mathrm{weight}}$ are listed in Table~\ref{tab:params_ppl}.

The input parameters required for the algorithm described in Section~\ref{sec:mat&methods:parametrization} were optimized manually to match the quasi-static pressure-volume points of the patient. 
The time constants $\tau$ of all terminal units were randomly chosen from a quasi-hyperbolic distribution, i.e., $\tau\in\frac{T}{\mathrm{unif}[0,1]}$, with the input parameter $T$, and $\mathrm{unif}[0,1]$ describing uniformly distributed stochastic values between 0 and 1~\cite{Bates2002}. 
We adapted the values for $\Delta P_{\mathrm{max-min}}$ and $\Delta P_{\mathrm{op-cl}}$ following the magnitude of variables with similar meaning in other models~\cite{Massa2008}. Further, the clinically applied and measured pressures entering the model calibration are provided in Table~\ref{tab:vent_params}.

\begin{figure*}[ht]
    \begin{center}
    \includegraphics[trim=0 0 0 20,clip,width=0.5\textwidth]{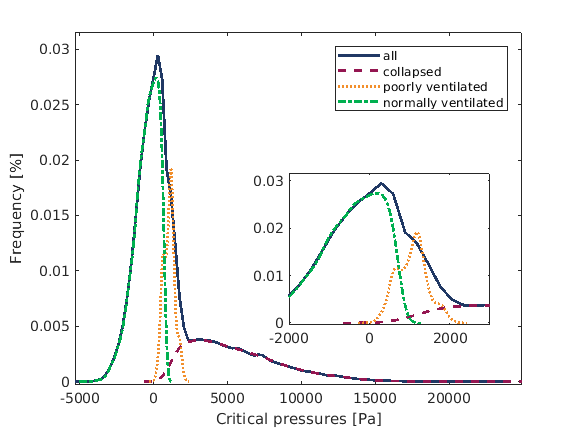}
    \caption{The kernel density functions of the $P_{\mathrm{op,crit,max}}$ of all terminal units and the subgroups of collapsed (red dashed), poorly-aerated (orange dotted), and normally ventilated (green dashed/dotted) terminal units; $P_{\mathrm{op,crit,min}}$, $P_{\mathrm{cl,crit,min}}$, and $P_{\mathrm{cl,crit,max}}$ behave accordingly.}
    \label{fig:critpress}
    \end{center}
\end{figure*}

The $P$-$V_{\mathrm{tot}}$ relations resulting from the input parameter values ultimately chosen and provided in Table~\ref{tab:input_params} are shown in Figure~\ref{fig:fit_lowflow}, together with the related clinical data curves. \textcolor{black}{Note, that for the normal breath cycle the matching is limited to the quasi-static points along the cycle, i.e., to end-inspiration and end-expiration (orange crosses in Figure~8). As a consequence of the assumptions of the parametrization algorithm, there are no resistive effects included in the simplified model underlying the calibration algorithm. Thus, periods of high flow such as in the beginning of inspiration and expiration in a normal breath cycle can not be reproduced adequately.}
Figure~\ref{fig:critpress} depicts the $P_{\mathrm{op,crit,max}}$ of all terminal units (also structured by pathological types), emerging from the algorithm and the given input parameters. $P_{\mathrm{op,crit,min}}$, $P_{\mathrm{cl,crit,min}}$, and $P_{\mathrm{cl,crit,max}}$ behave accordingly.

\subsection{Results}
\label{sec:results}

\begin{figure*}[ht]
\centering
\begin{minipage}{.45\textwidth}
  \centering
  \includegraphics[trim=110 100 140 80,clip,width=1.0\textwidth]{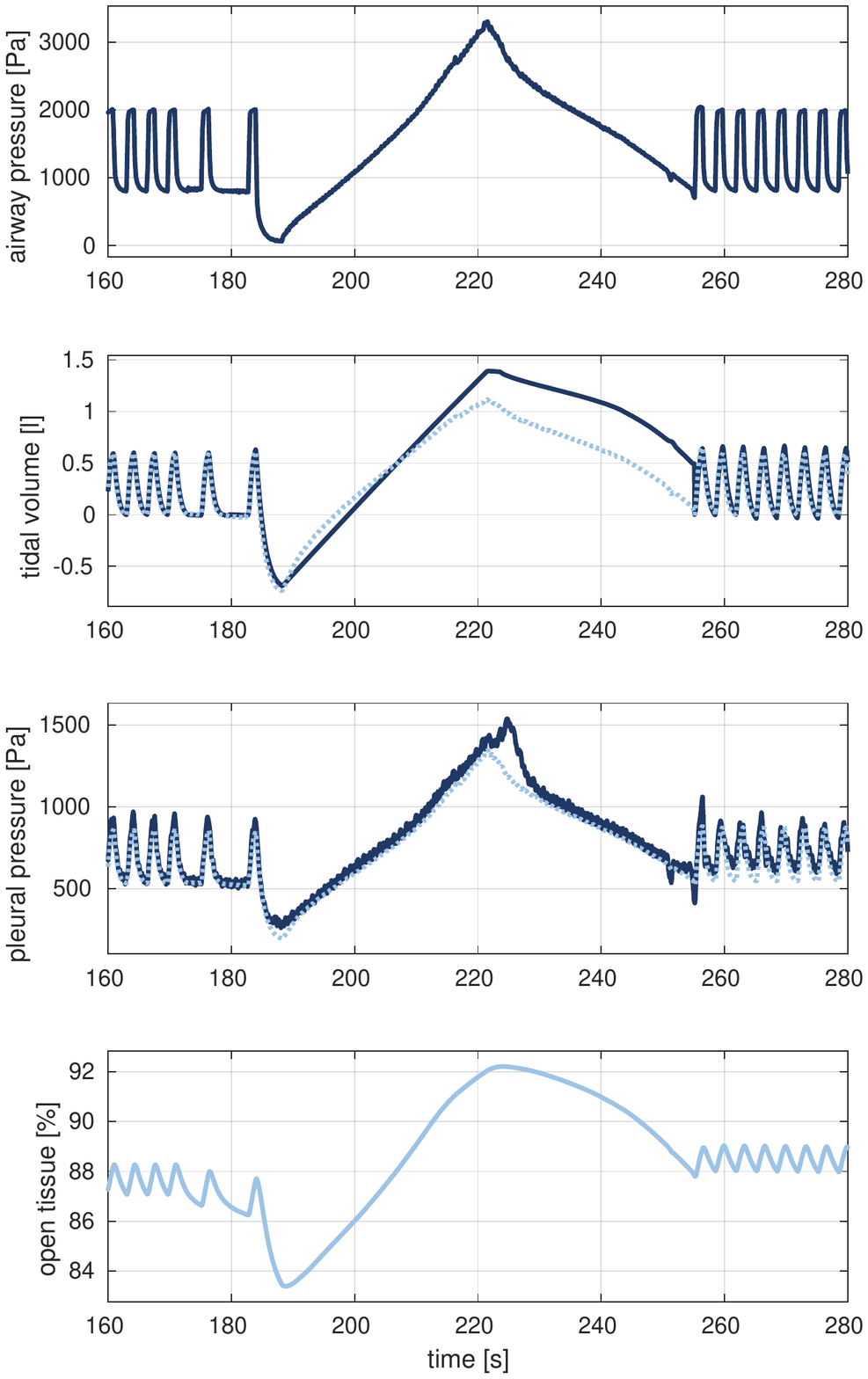}
  \captionof{figure}{Simulation results of the patient-specific computational lung model for a clinically applied airway pressure~(top) consisting of normal ventilation and a quasi-static inflation maneuver starting at ZEEP: From top to bottom, the tidal volume and pleural pressure produced by the model~(dotted light blue) and measured int the clinic~(solid dark blue), and the percentage of open tissue in the whole model~(bottom).}
  \label{fig:results_lf2}
\end{minipage}\qquad
\begin{minipage}{.45\textwidth}
  \centering
  \includegraphics[trim=110 100 140 80,clip,width=1.0\textwidth]{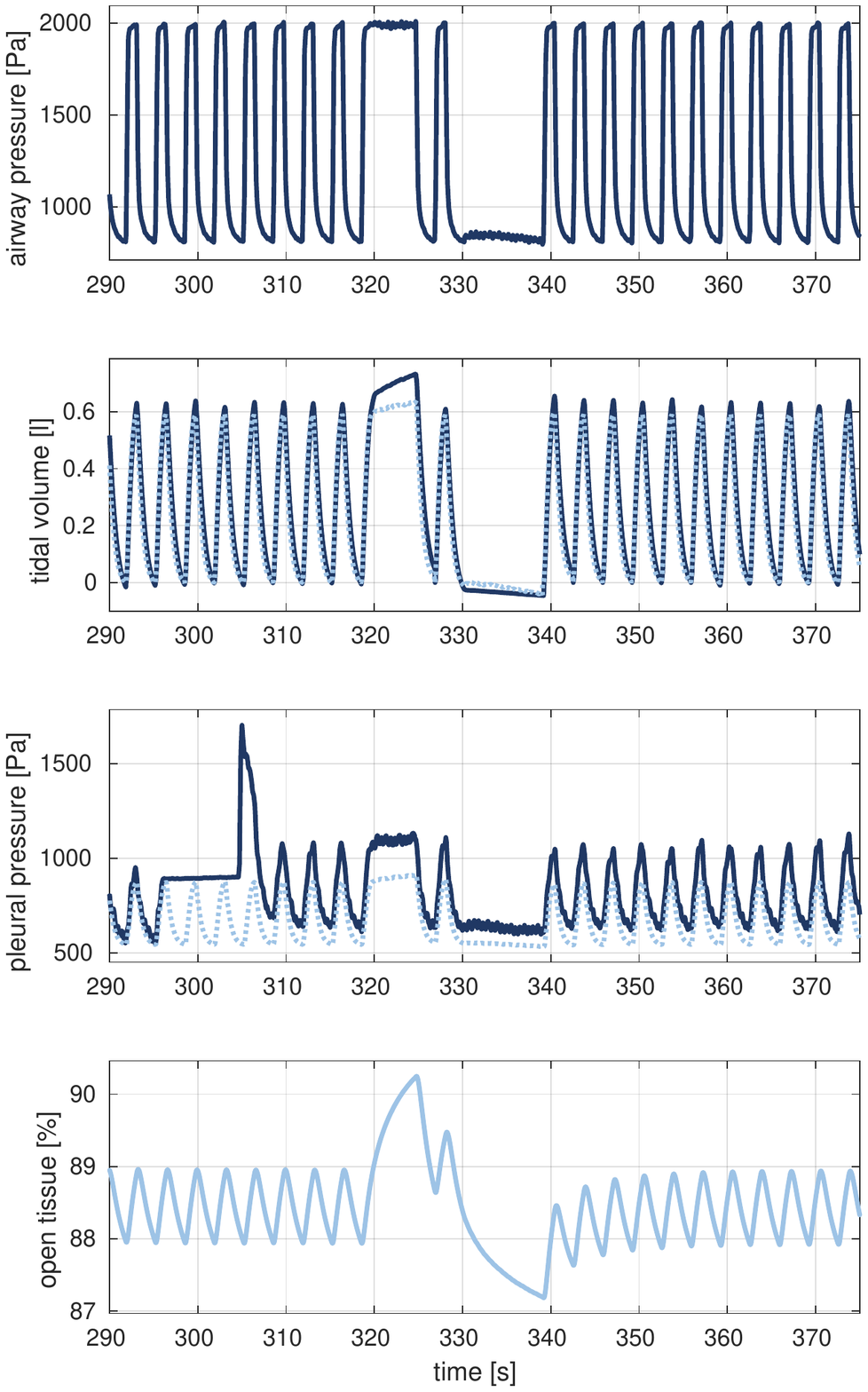}
  \captionof{figure}{Simulation results of the patient-specific computational lung model for a clinically applied airway pressure~(top) consisting of normal ventilation and an inspiratory hold maneuver without occlusion of the airflow: From top to bottom, the tidal volume and pleural pressure produced by the model~(dotted light blue) and measured int the clinic~(solid dark blue), and the percentage of open tissue in the whole model~(bottom).}
  \label{fig:results_insphold_occl0}
\end{minipage}
\end{figure*}

\begin{figure*}[ht]
\centering
\begin{minipage}{.45\textwidth}
  \centering
  \includegraphics[trim=110 100 140 80,clip,width=1.0\textwidth]{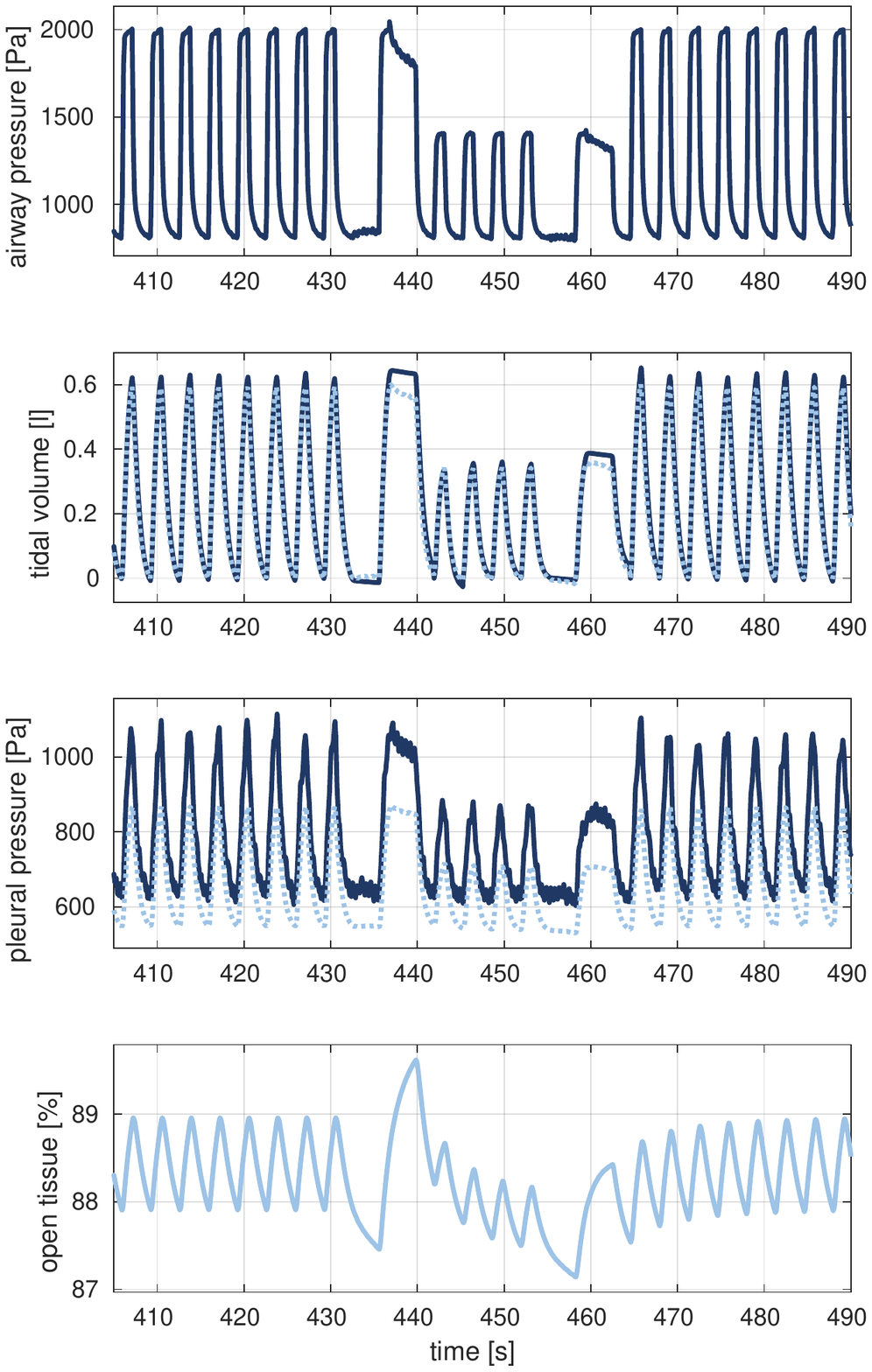}
  \captionof{figure}{Simulation results of the patient-specific computational lung model for a clinically applied airway pressure~(top) consisting of normal ventilation and two inspiratory hold maneuvers with occluded airflow combined with temporarily halved driving pressure: From top to bottom, the tidal volume and pleural pressure produced by the model~(dotted light blue) and measured int the clinic~(solid dark blue), and the percentage of open tissue in the whole model~(bottom).}
  \label{fig:results_insphold_occl1_deltap}
\end{minipage}\qquad
\begin{minipage}{.45\textwidth}
  \centering
  \includegraphics[trim=110 100 140 80,clip,width=1.0\textwidth]{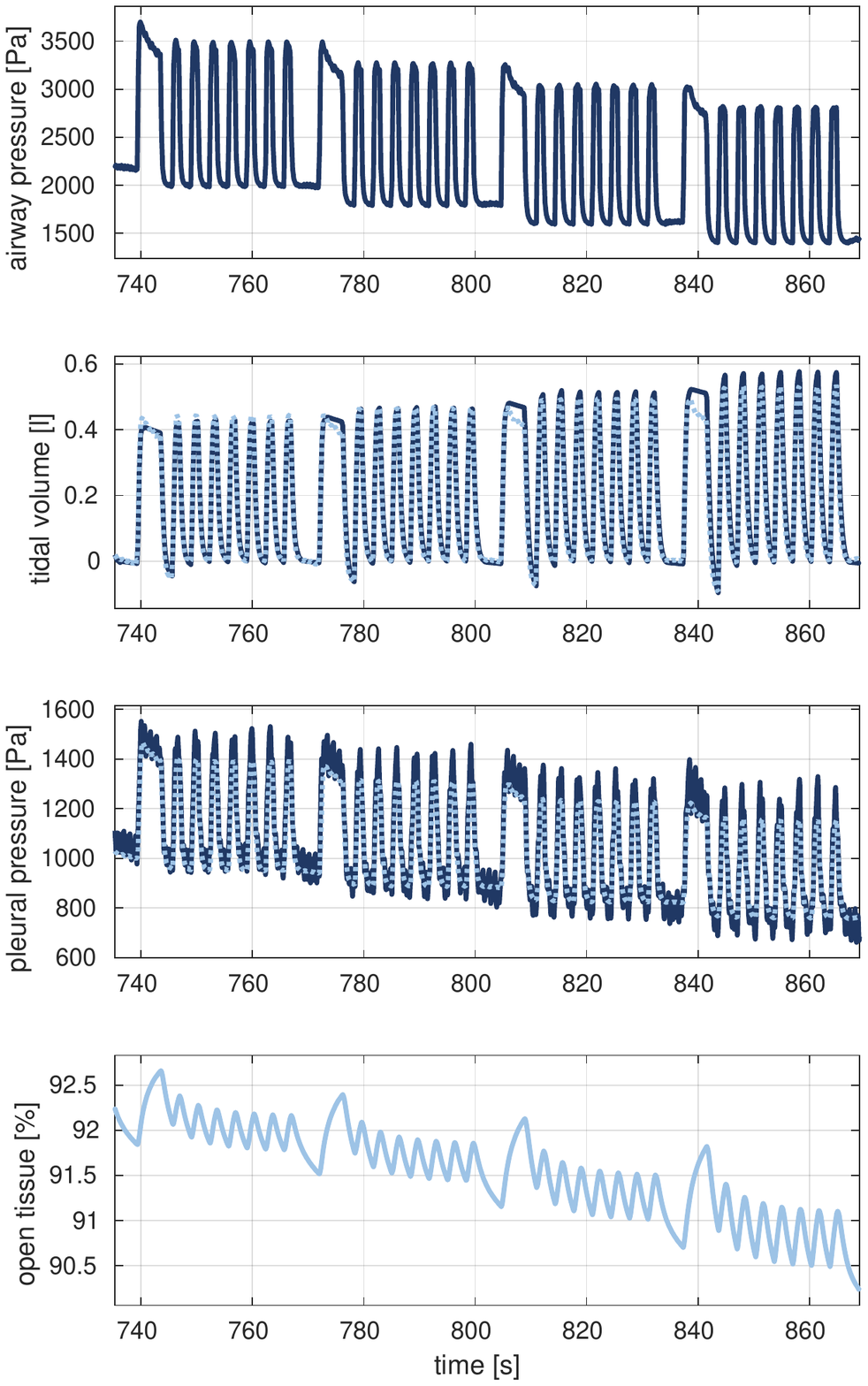}
  \captionof{figure}{Simulation results of the patient-specific computational lung model for a clinically applied airway pressure~(top) consisting of normal ventilation with a stepwise decremented PEEP, where each step was initiated by an inspiratory hold maneuver with occluded airflow: From top to bottom, the tidal volume and pleural pressure produced by the model~(dotted light blue) and measured int the clinic~(solid dark blue), and the percentage of open tissue in the whole model~(bottom).}
  \label{fig:results_decrpeep}
\end{minipage}
\end{figure*}

\paragraph{Simulation protocol}
Using the generated model tailored to a patient's lung we simulated 30 minutes of mechanical ventilation of the patient. We applied the airway pressure obtained from the clinical measurements at the airway opening of the simulation model for this purpose. In order to validate our model, we are presenting several time ranges, including various maneuvers, from this simulation study. Note that we are in this case excluding the quasi-static inflation maneuver used for calibration of the model in Section~\ref{sec:mat&methods:modelconcrete} to not simply reproduce the data the model was fed with.\\
Apart from cycles of normal ventilation, the ranges extracted from the study contain the following maneuvers:
\begin{enumerate}
    \item A quasi-static inflation maneuver that covers a wide pressure range of the patient's respiratory system (Figure~\ref{fig:results_lf2}). At a starting level at zero end-expiratory pressure~(ZEEP) and reaching a peak airway pressure of 33~mbar, the maneuver captures a broader pressure range than the quasi-static inflation maneuver used for model calibration,
    \item An inspiratory hold maneuver where the phase of inspiration was prolonged by keeping the driving pressure constant for 6~s (Figure~\ref{fig:results_insphold_occl0}), allowing an additional volume increase in the lung, inter alia, due to RD; this maneuver is followed by an expiratory hold maneuver, where the flow was kept close to zero,
    \item An inspiratory hold maneuver with occluded tracheal airflow, followed by normal ventilation cycles with half the original driving pressure and another inspiratory hold maneuver with flow occlusion at reduced driving pressure (Figure~\ref{fig:results_insphold_occl1_deltap}), and
    \item A decremental PEEP trial, where the PEEP level is gradually reduced from an elevated level, each time started by an inspiratory and ended by an expiratory hold maneuver with flow occlusion (Figure~\ref{fig:results_decrpeep}).
\end{enumerate}
The time in Figures~\ref{fig:results_lf2}~-~\ref{fig:results_decrpeep} indicates the chronological sequence of the extracted ventilation periods.\\

\paragraph{Global mechanics}
Figures~\ref{fig:results_lf2}~-~\ref{fig:results_decrpeep} illustrate the computationally predicted global response of the lung model to the applied airway pressure profile~(top). From top to bottom, they depict the tidal volume and the pleural pressure of the simulation results and the clinical measurements. Further, the percentage of open reference volume in the whole lung model~(bottom), i.e.,~summed over all terminal units, indicates the occurrence of RD, when it changes. 

\noindent\textbf{Ventilation quantities} \quad In overall terms, the simulation results were close to the clinical data. Along normal ventilation, both the tidal volume and pleural pressure followed the shape and amplitude of the measured breath cycles well, be it for original or halved driving pressure (Figure~\ref{fig:results_insphold_occl1_deltap}). 
The same was true of the prolonged periods of expiration, and especially when the airway pressure is decreased toward ZEEP, prior to the quasi-static inflation maneuver (Figure~\ref{fig:results_lf2}). Therefore, the model also captures the volume course in the lower pressure range very well. \\
Note that the clinical volume curve calculated from the measured airflow exhibited an unpredictible baseline drift~---~potentially due to sensor disturbances caused by the humidified respiratory air~--- resulting in a disinctively varying end-expiratory volume level. As a result, we compared the simulation and clinical data in a tidal manner only. However, the drift in the quasi-static inflation maneuver was notably strong~(Figures~\ref{fig:results_lf2}) when there was a gain in the measured volume of about 0.4~l compared to the volume at PEEP before the maneuver, without causing a similar tendency in the measurements of the actually directly coupled pleural pressure. The close agreement between the clinical and the simulated pleural pressures along the maneuver justifies the validity of our model. Further, at the same pleural pressures during the quasi-static inflation maneuver ($t = 203.5~\mathrm{s}$ and $t = 250~\mathrm{s}$) the volume values should also be similar, which is the case for the simulation.

Evaluating the pleural pressure in more detail, the simulation values closely resemble the clinical data. Note that single swallowing actions by the patient caused temporary disturbances in the measurements or failures of the measurement sensor (e.g., see Figures~\ref{fig:results_lf2} at $t = 225~\mathrm{s}$ or \ref{fig:results_insphold_occl0} at $t = 305~\mathrm{s}$, and \ref{fig:results_insphold_occl0} at $t = 295~\mathrm{s}$, respectively)~\cite{Dornhorst1952}.
The pleural pressure base level sometimes changed after such events (e.g., starting in Figure~\ref{fig:results_insphold_occl0}, enduring in Figure~\ref{fig:results_insphold_occl1_deltap}, and again reset in Figure~\ref{fig:results_decrpeep}). The variation in pleural pressure, however, can still serve for comparison despite a varying base level of the pressure~\cite{Cammarota2023}, and it does match well for our simulation results and the measurements.\\
\noindent\textbf{RD dynamics} \quad The accumulated behavior of alveolar RD can be seen in Figures~\ref{fig:results_lf2}~-~\ref{fig:results_decrpeep} by the percentage of open tissue volume. With the overall open stress-free reference volume of the model of $3.25~\mathrm{l}$ which theoretically gives a gas volume of about $3.69~\mathrm{l}$ at $P_\mathrm{tp,PEEP}$~(global) when solving Eq.~\eqref{eq:Ogden}, one percent of open tissue corresponds to an additional gas volume of approximately $37~\mathrm{ml}$ at $P_\mathrm{tp,PEEP}$~(global). We see continuous intratidal RD, and the single maneuvers have remarkable effects on the degree of recruited volume. As expected, the degree of RD in the model is strongly influenced by the level of airway pressure~(Figure~\ref{fig:results_decrpeep}). The effect of time dependence is especially obvious during the longer periods of expiration and inspiration, e.g., prior to the quasi-static inflation maneuver, or during the inspiratory and expiratory hold maneuvers (Figure~\ref{fig:results_insphold_occl0}), when there is ongoing recruitment or derecruitment, respectively. 

Moreover, stable recruitment of tissue after the quasi-static inflation maneuver is evident by the $\sim$~1\%~increase in the oscillating open tissue compared to the time before the maneuver (Figure~\ref{fig:results_lf2}). However, we see that the model slightly underestimates the tidal volume after the inflation maneuver, i.e., the volume gain is obviously not sufficient. A similar effect was observed in the inspiratory hold maneuvers (Figures~\ref{fig:results_insphold_occl0} and~\ref{fig:results_insphold_occl1_deltap}) and at the end of the decremental PEEP trial in Figure~\ref{fig:results_decrpeep}. 
Hence, although the compliance increases due to the opening of tissue, the overall permanently recruiting volume in the model is insufficient.

Another interesting aspect is that in the decremental PEEP trial the tidal RD exhibited a smaller amplitude in the higher pressure regions than in the lower pressure regions (Figure~\ref{fig:results_decrpeep}). In addition, reducing the driving pressure apparently also decreases the degree of tidal RD (Figure~\ref{fig:results_insphold_occl1_deltap}).

\paragraph{Local mechanics}

\begin{figure*}[ht]
    \begin{center}
    \includegraphics[trim=0 210 70 0,clip,width=0.7\textwidth]{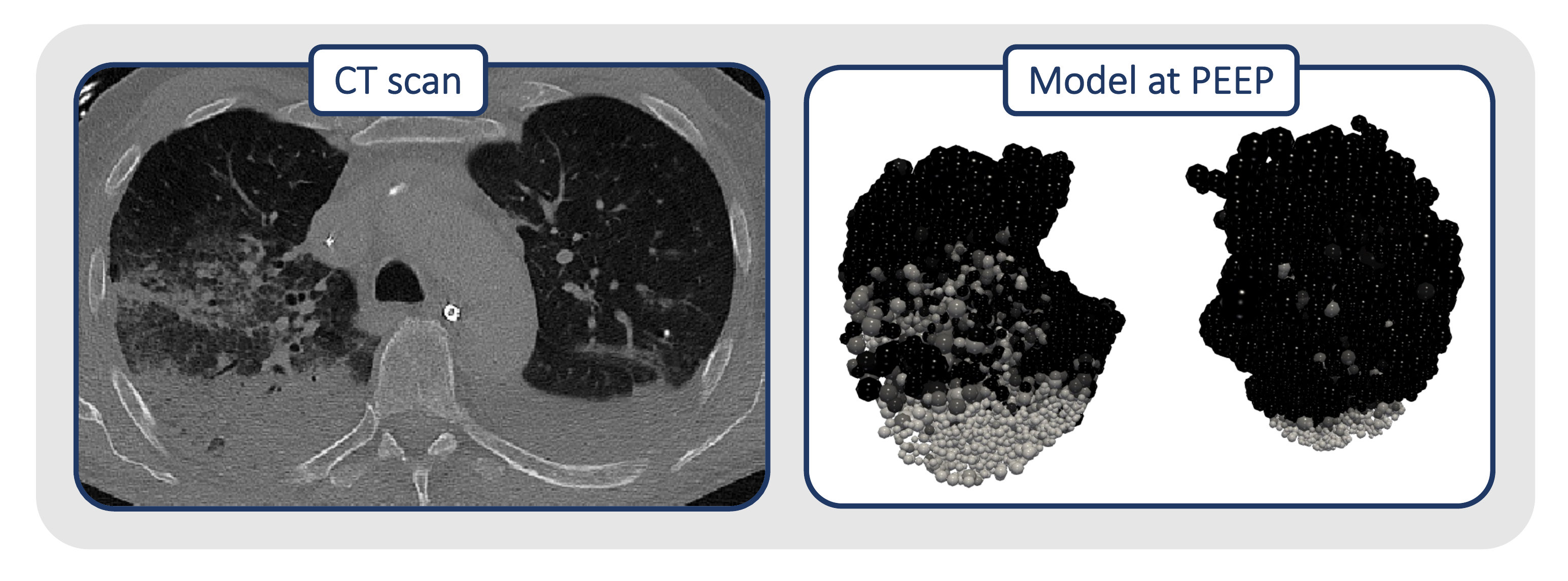}
    \caption{Comparison of the CT scan and the opening proportion of terminal units gradually from fully closed (grey) to fully open (black), in the patient-specific lung model at PEEP ($t = 31.6~\mathrm{s}$).}
    \label{fig:results_comp_ct_peep}
    \end{center}
\end{figure*}

\begin{figure}[ht]
    \centering
    \includegraphics[page=2,trim=0 50 0 0,clip,width=0.7\textwidth]{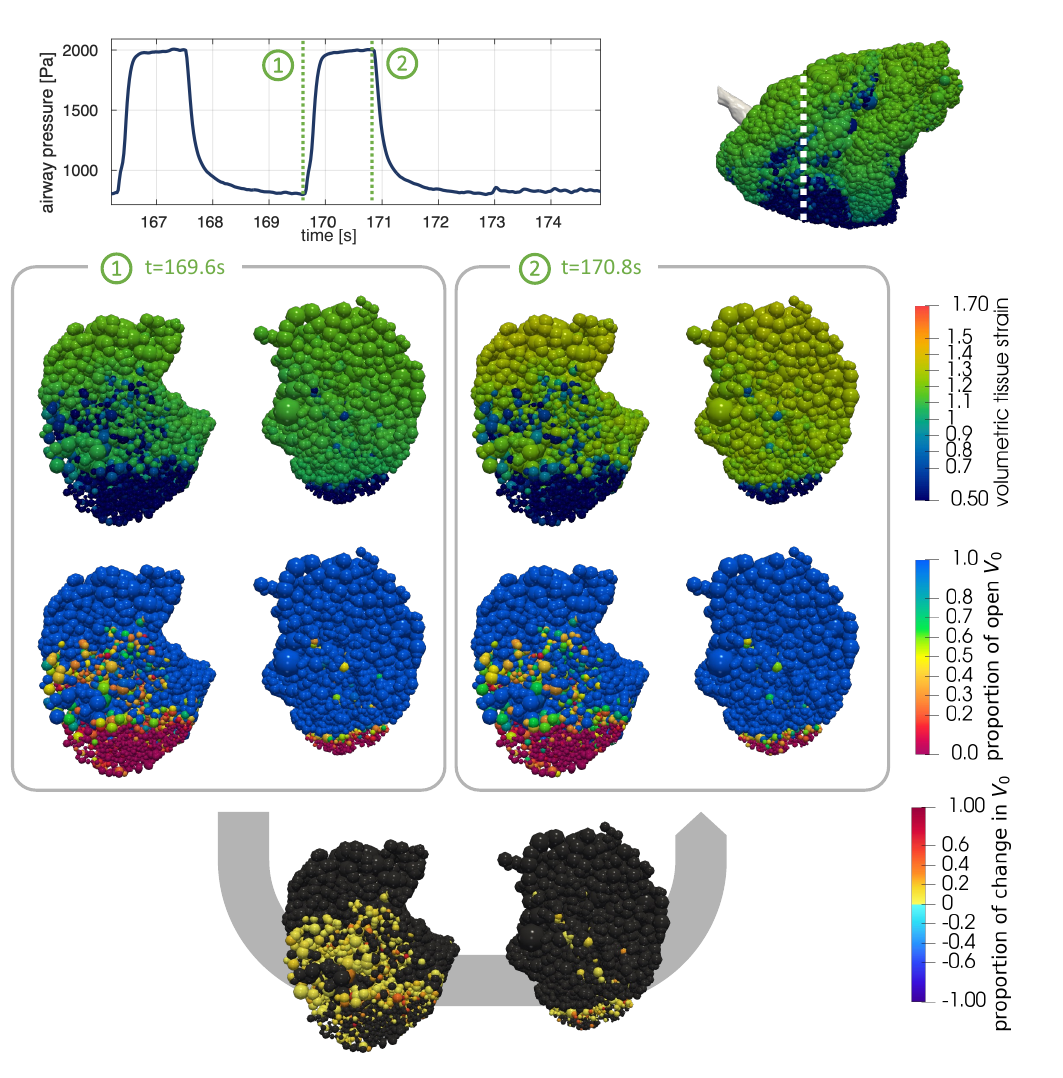}
    \caption{Local model behavior during a normal breath cycle at end-expiration ($t = 169.9~\mathrm{s}$,~\textcircled{1}) and end-inspiration ($t = 170.8\mathrm{s}$,~\textcircled{2}); from top to bottom: volumetric tissue strain in the terminal units, their recruitment state indicated by the proportion of open reference volume~$V_{0}$ (from fully closed~=~0 to fully open~=~1), and the difference in recruitment between the specified time points (i.e., the proportional change of open reference volume~$V_{0}$); view: axial cut through the supine lung model.}
    \label{fig:results_local_normal}
\end{figure}

Figure~\ref{fig:results_comp_ct_peep} illustrates a qualitative comparison between the CT scan used for model generation and parametrization and the ventilation of the terminal units at PEEP ($t = 31.6~\mathrm{s}$), where their recruitment states (from closed to open) are mapped to the range of grey values shown in the medical image. This result is intended as a proof of concept for the image- and ventilation-based model parametrization described because it is illustrative of our model's ability to capture the ventilation state of the patient's lung at PEEP accordingly. The actual values for local strains and recruitment states of the model at PEEP are shown in Figure~\ref{fig:results_local_normal}. 

\begin{figure}[ht]
    \centering
    \includegraphics[page=2,trim=0 20 15 0,clip,width=1.0\textwidth]{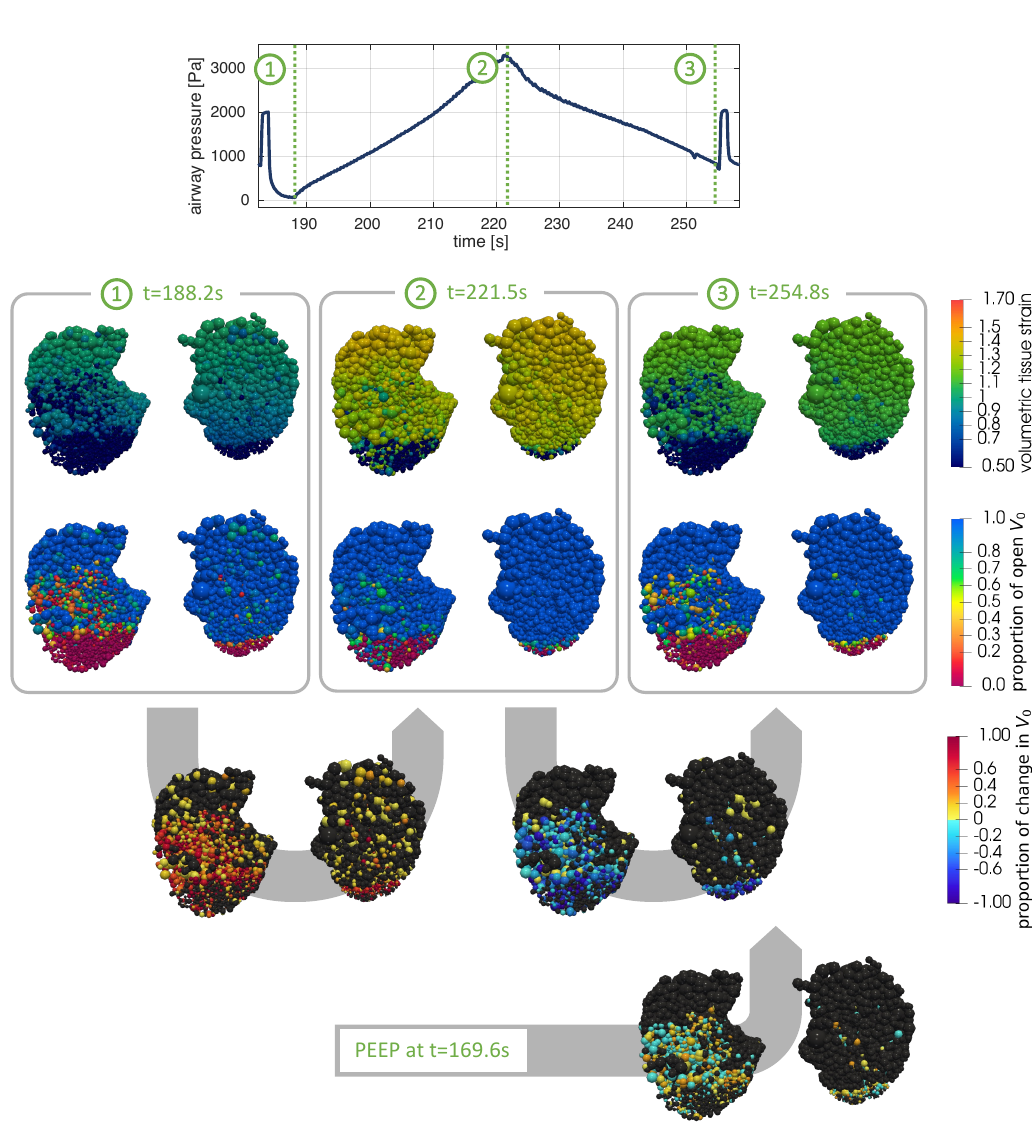}
    \caption{Local model behavior along the quasi-static inflation maneuver at the start ($t = 188.2~\mathrm{s}$,~\textcircled{1}), at the peak ($t = 221.5~\mathrm{s}$,~\textcircled{2}) and at the end ($t = 254.8~\mathrm{s}$,~\textcircled{3}) of the maneuver; from top to bottom: volumetric tissue strain in the terminal units, their recruitment state indicated by the proportion of open reference volume~$V_{0}$ (from fully closed~=~0 to fully open~=~1), and the difference in recruitment between the specified time points (i.e., the proportional change of open reference volume~$V_{0}$); the difference in recruitment at PEEP8 before and after the maneuver is evaluated between $t = 169.6~\mathrm{s}$ and $t = 254.8~\mathrm{s}$~(bottom); view: axial cut through the supine lung model.}
    \label{fig:results_local_lf2}
\end{figure}

\begin{figure}[ht]
    \centering
    \includegraphics[page=1,trim=0 110 15 0,clip,width=1.0\textwidth]{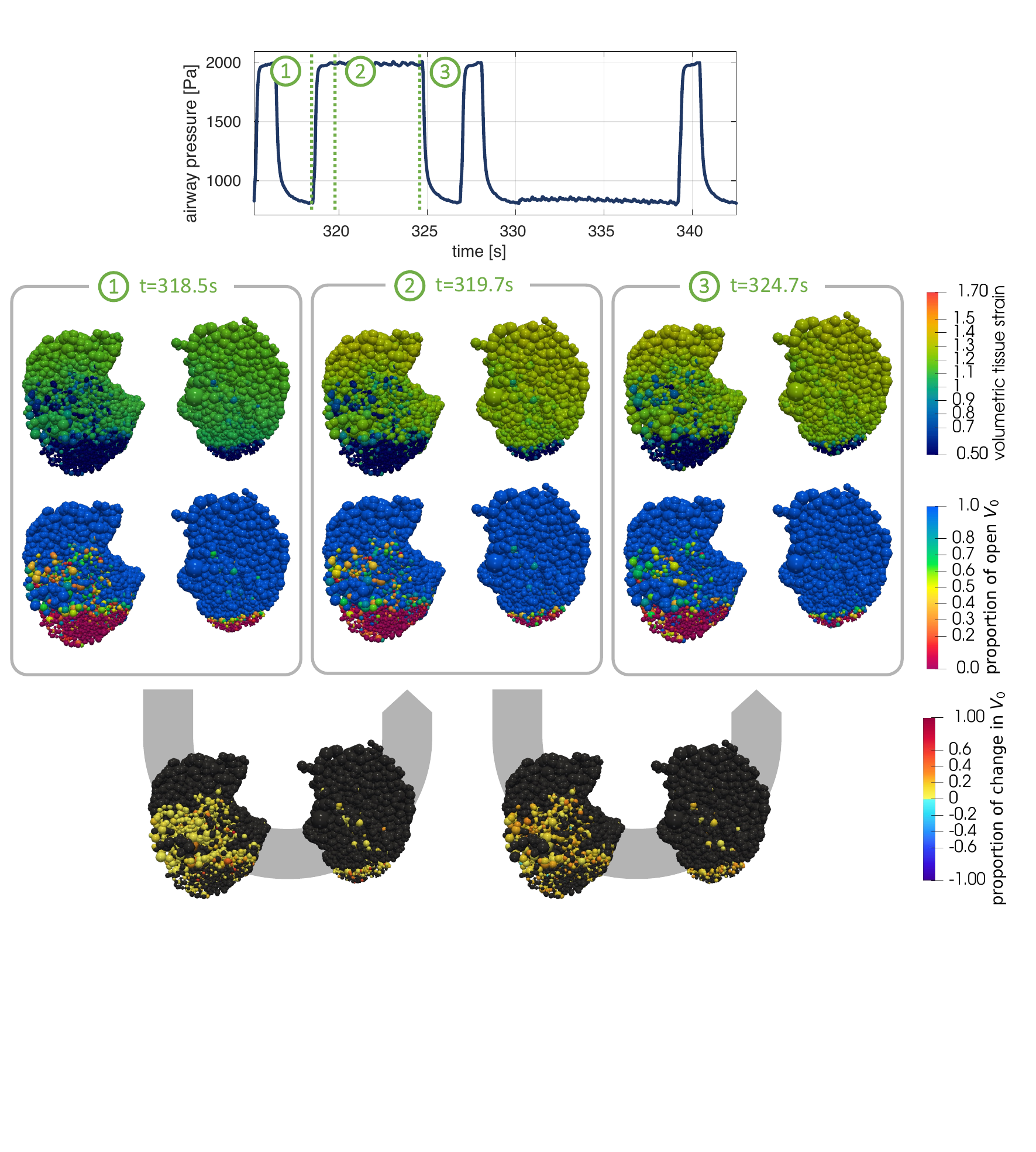}
    \caption{Local model behavior along the inspiration of an inspiratory hold maneuver at end-expiration before the maneuver onset ($t = 318.5~\mathrm{s}$,~\textcircled{1}), after the inspiration time of a normal breath cycle ($t = 319.7~\mathrm{s}$,~\textcircled{2}) and at end-inspiration ($t = 324.7~\mathrm{s}$,~\textcircled{3}); from top to bottom: volumetric tissue strain in the terminal units, their recruitment state indicated by the proportion of open reference volume~$V_{0}$ (from fully closed~=~0 to fully open~=~1), and the difference in recruitment between the specified time points (i.e., the proportional change of open reference volume~$V_{0}$); view: axial cut through the supine lung model.}
    \label{fig:results_local_inspholdoccl0_insp}
\end{figure}

\begin{figure}[ht]
    \centering
    \includegraphics[page=2,trim=0 70 15 0,clip,width=1.0\textwidth]{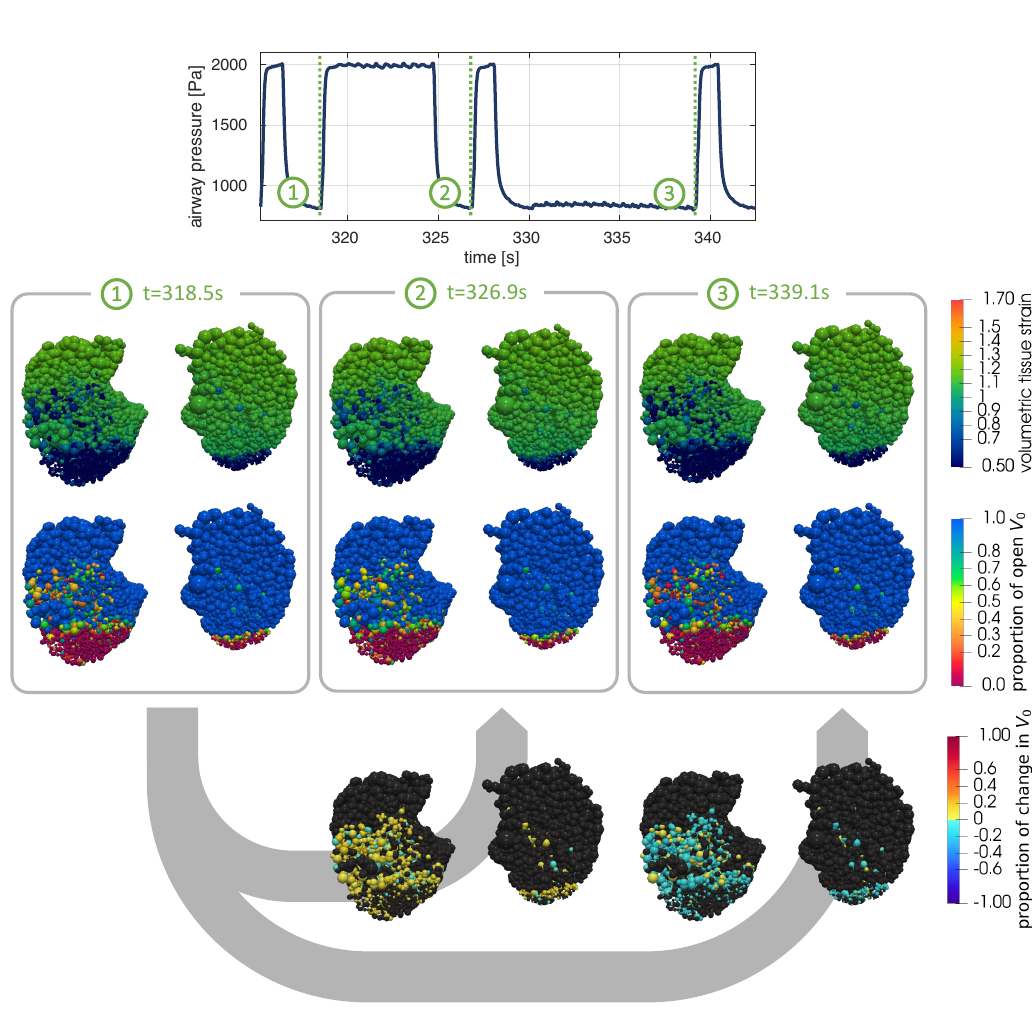}
    \caption{Local model behavior at points of end-expiration before ($t = 318.5~\mathrm{s}$,~\textcircled{1}) and after ($t = 326.9~\mathrm{s}$,~\textcircled{2}) an inspiratory hold, and at the end of an appended expiratory hold ($t = 324.7~\mathrm{s}$,~\textcircled{3}); from top to bottom: volumetric tissue strain in the terminal units, their recruitment state indicated by the proportion of open reference volume~$V_{0}$ (from fully closed~=~0 to fully open~=~1), and the difference in recruitment between the specified time points (i.e., the proportional change of open reference volume~$V_{0}$); view: axial cut through the supine lung model.}
    \label{fig:results_local_inspholdoccl0_exp}
\end{figure}

To give an impression of the regional model behavior for specific ventilation maneuvers, Figures~\ref{fig:results_local_normal}~-~\ref{fig:results_local_inspholdoccl0_exp} illustrate the local volumetric strains, the current recruitment state (proportion of open reference volume), and the difference in recruitment between specific time points, all for (i)~a normal breath (Figure~\ref{fig:results_local_normal}), (ii)~along the quasi-static inflation maneuver (Figure~\ref{fig:results_local_lf2}) not used for parametrization, and for periods of (iii)~inspiratory and (iv)~expiratory hold (Figures~\ref{fig:results_local_inspholdoccl0_insp} and~\ref{fig:results_local_inspholdoccl0_exp}, respectively).
We calculated the volumetric strain of a terminal unit by $\epsilon_{vol} = (V + V_{\mathrm{tissue}}) / (V_{\mathrm{0,max}} + V_{\mathrm{tissue}})$. Note that the straining of a terminal unit refers to its stress-free state, and not to the functional residual capacity~(FRC) or the end-expiratory lung volume~(EELV), as it is usually the case in the medical community. We then have the opportunity to evaluate the absolute strain, and not the strain from a specific state. 
The proportion of open reference volume,~$k_{\mathrm{open}}(t)$, in a terminal unit was determined according to $k_{\mathrm{open}}(t) = (V_{\mathrm{0}}(P,t) - V_{\mathrm{0,min}}) / (V_{\mathrm{0,max}} - V_{\mathrm{0,min}})$. To evaluate the amount of RD between two states $t_i$ and $t_{i+1}$, we computed the differences in the proportion of open reference volume by $k_{\mathrm{open}}(t_{i+1}) - k_{\mathrm{open}}(t_i)$, such that negative values indicate the closing proportion of $V_{\mathrm{0,max}}$ of a terminal unit, and positive values the opening proportion, respectively. Black colored terminal units do not exhibit any change in reference volume between the specified time points. 


As expected, we see that the heterogeneity in straining is directly linked to the recruitment state of a terminal unit since (partially) closed terminal units generally react in a stiffer manner (see the volumetric strains and proportion of open reference volumes in Figures~\ref{fig:results_local_normal}~-~\ref{fig:results_local_inspholdoccl0_exp}).
Note that we obtain a homogeneous strain behaviour across the lung regionally only influenced by the gravitational pressure gradient if we determine the relative strain based on the currently open tissue by $\epsilon_{vol} = (V + V_{\mathrm{tissue}}) / (V_{\mathrm{0}}(P,t) + V_{\mathrm{tissue}})$, i.e., although the overall strain in a terminal unit may appear low, the open tissue of the terminal unit can experience higher strains.

As would be expected given the assumptions in model calibration, the tidal recruitment in normal ventilation occurs predominantly in partially collapsed terminal units or at the edges of atelectases, and to a differing degree (Figure~\ref{fig:results_local_normal}, bottom). It is further evident that the regional pattern of tidal RD greatly resembles from breath to breath in normal ventilation, in accordance with the mathematical assumptions we made in Section~\ref{sec:mat&methods:parametrization}. 

Along the quasi-static inflation maneuver, a high amount of RD occurs spread across the lung (Figure~\ref{fig:results_local_lf2}), leading to several stably recruiting terminal units, which we previously recognized already in the global RD curves (Figure~\ref{fig:results_lf2}). After significant derecruitment occurring in terminal units all over the lung model when lowering the airway pressure to ZEEP, the model exhibits a high amount of (re-)opening along the inspiratory branch of the maneuver. This even continues very slightly for a few terminal units during expiration from peak pressure back to PEEP (see yellow terminal units in comparison of the proportion of change in reference volume between $t = 221.5~\mathrm{s}$ and $t = 254.8~\mathrm{s}$ in Figure~\ref{fig:results_local_lf2}). The vast majority of terminal units transitioning during expiration, however, shows a closing behavior (see blue colored terminal units). 
When evaluating the net impact of the quasi-static inflation between PEEP at the end of the maneuver ($t = 254.8~\mathrm{s}$) and the last regular PEEP in normal ventilation before the maneuver ($t = 169.9~\mathrm{s}$), the recruitment prevails (see yellow-red colored terminal units in Figure~\ref{fig:results_local_lf2}, bottom).

Interestingly, we also observe small derecruitment in several terminal units, again close to zero (see cyan colored terminal units in Figure~\ref{fig:results_local_lf2}, bottom). We traced this phenomenon back to the volume-dependent pleural pressure, $P_{\mathrm{pl}}$: The permanent recruitment of terminal units during the quasi-static inflation increases the total gas volume of the model at PEEP, which again causes a small raise in $P_{\mathrm{pl}}$. The thereby slightly reduced pressure difference across the terminal units leads to some derecruitment, especially in terminal units that are open at PEEP8, but tend to close due to a critical closing pressure close to $P_{\mathrm{tp,PEEP}}$. In total, however, this effect is very small and thus not observable in the global RD curve (Figure~\ref{fig:results_lf2}, bottom). 
For a more detailed view, the RD between the specified points in time is quantified for all affected terminal units over the lung height in Figure~\ref{fig:results_local_RD} in the appendix.

The effect of time dependence in RD is particularly evident along the inspiratory hold maneuver in Figure~\ref{fig:results_local_inspholdoccl0_insp}. At $t = 319.7~\mathrm{s}$, the time when the inspiration of a normal breath cycle would end, the model exhibits a similar RD pattern as in normal ventilation (Figure~\ref{fig:results_local_inspholdoccl0_insp}, bottom left, and Figure~\ref{fig:results_local_normal}, bottom). The subsequent prolonged inspiration triggers a considerable amount of additional recruitment until $t = 319.7~\mathrm{s}$ (Figure~\ref{fig:results_local_inspholdoccl0_insp}, bottom right), which appeared already from the global RD curve in Figure~\ref{fig:results_insphold_occl0}. 

In Figure~\ref{fig:results_local_inspholdoccl0_exp}~(bottom left), we see that many terminal units continue to have a higher amount of recruited volume when returning to PEEP at $t = 326.9~\mathrm{s}$ compared to the end-expiration before the maneuver at $t = 318.5~\mathrm{s}$. Here again, we identify a slight and at first counterintuitive derecruitment in some terminal units similar to the observations along the quasi-static inflation maneuver and again caused by the volume-dependent pleural pressure (for more details see Figure~\ref{fig:results_local_RD}). 
The reverse effect appears at the end of an endured period of expiration at $t = 339.1~\mathrm{s}$ (Figure~\ref{fig:results_local_inspholdoccl0_exp}, bottom right, and Figure~\ref{fig:results_local_RD}). Compared to $t = 318.5~\mathrm{s}$, the majority of terminal units experiences derecruitment at the end of the applied combination of inspiratory and expiratory hold maneuvers, but there are some terminal units slightly opening up.

\section{Discussion and conclusion}
\label{sec:discussion}

In current clinical practice, ventilation management is performed in a generic manner, inter alia, according to the patient's estimated height and ideal body weight~\cite{Ranieri2012}. Such limited attempts to adapt the clinical treatment to a specific patient lack local insight into the patient's lung and disregard the specific pathology, both of which are important aspects that might have a crucial impact on the minimization of VILI, and eventually the reduction of mortality due to ARDS~\cite{Kollisch-Singule2018,ARDS2000}. 

In this study, we presented a model for lung tissue incorporating a novel approach accounting for the harmful phenomena of OD and repetitive RD, both of which are widely recognized as contributors to the pathogenesis of VILI. The mechanism of RD is implemented by the pressure- and time-dependent variation of stress-free reference volume. By means of the underlying viscoelastic model components, the approach involves time dependence due to both collapse and opening, and tissue resistance. OD is captured by the open reference volume subjected to straining. In principle, overdistension and collapse can occur at the same time in a tissue component.

Concurrently, we proposed a method to apply the newly introduced RD model multi-compartmentally in the framework of an anatomically accurate computational lung model and to tailor the model to specific patients. The algorithmic parametrization of each of the numerous tissue components in the lung model described herein considers the local pathology of the injured organ in the manner indicated by medical imaging. Therefore, a terminal unit reproduces the degree of RD condition of the modelled region as specified by the CT scan, i.e., fully collapsed, poorly aerated (and thus gradually recruited), or normally ventilated at the pressure level of the lung in the recorded image. In order to also capture the pressure-volume behavior of the lung beyond this single, end-expiratory state, the algorithm input parameters are optimized such that the lung model matches the patient's global respiratory mechanics as observed in clinical measurements. The influence of the chest wall on the lung is accounted for by a customized lung volume-dependent pleural pressure boundary condition acting outside the terminal units.

As an application, we deployed the full modeling concept to an example patient suffering from ARDS and receiving mechanical ventilation in the intensive care unit. In simulations of various clinical airway pressure profiles, we tested the model response and compared the resulting tidal volume and pleural pressure to evaluate the predictive capability of the model. Both quantities show very reasonable results and closely follow the measured data (Figures~\ref{fig:results_lf2}~--~\ref{fig:results_decrpeep}). The alveolar RD model elements mimicked repetitive intra-tidal RD, stable recruitment after elevated pressure levels, and transient opening or closing when experiencing changes in the ventilation mode, or along constant pressure profiles. In addition, the parametrization of the exemplary lung model appears plausible when comparing the CT scan to the local model behavior. Moreover, we observe the characteristic of non-normally distributed but scattered critical pressures in the resulting model parameters (Figure~\ref{fig:critpress}), which was also observed experimentally~\cite{Fardin2021,Albert2009}. 

This novel modeling concept demonstrates a variety of promising and significant aspects with regard to future patient-specific treatments:

Firstly, the design of the presented alveolar RD model and the method of customized parametrization in a full lung model will enable us to overcome the lack of knowledge and the inaccessibility of biophysical properties of the local lining liquid in air spaces, which emerged in our previous work~\cite{Geitner2022} and has hindered the predictive patient-specific modeling. It is still impossible to obtain the actual physical quantities crucial to pathological behavior in diseased regions of the lung. We are instead using information about the physical properties of lining fluid of deranged lung units intrinsically included in the CT scan by the gray values, e.g., indicating a collapse tendency of a region at the specific pressure state due to the composition of the present fluid. In the proposed model, this information is phenomenologically integrated into the RD parameters.

Secondly~---~to offer a glimpse into the future potential~---~, a more realistic reproduction of volumetric behavior and, thus, distribution of air in the lung can enable a more realistic estimation of the surface available for oxygen exchange. However, drawing any conclusions about blood oxygenation will of course require extending the present approach by an adequate model for gas exchange and lung perfusion.

Last but not least, we again stress the ability of the multi-compartment model presented to capture and evaluate two major contributors to VILI, and we further emphasize in this context especially the essential inclusion of time dependence in RD~\cite{Fardin2021}. Individually tailored to a patient's lung geometry, to the overall mechanics of the thoracic cage and to the lung pathologies~---~be it the elevated lung weight, or the local collapse tendencies and its dynamics~--- this model can deliver a measure of the injury potential of ventilation profiles with respect to a specific lung. By changing the reference volume, we can directly access the amount of repetitive reopening which produces harmful shear stresses at airway walls in real life \textcolor{black}{as well as the opening velocity}~\cite{Bilek2003,Kay2004}, and we can further detect and quantify excessive straining. \textcolor{black}{Previously developed approaches}~\cite{Hamlington2016,Mellenthin2019,Bates2020} can serve as estimation for volutrauma and atelectrauma in our model, but in a patient-specific and locally resolved manner. This direction will certainly require further research in multi-patient studies, which is currently ongoing, and a thorough investigation and definition of safe thresholds for the measures mentioned.

\textcolor{black}{To give an even broader perspective on the usability of the novel modeling approach, we outline~---~just in a nutshell~---~some potential ideas how such models can enhance mechanical ventilation. One straightforward and evident concept is to leverage such a model to provide additional insight and data concerning non-measurable quantities deep down in the lung during ventilation. Physicians could utilize this information to differentiate between cases and guide subsequent treatment decisions. Additionally, such (so far completely missing) data points could be very helpful when employing novel AI-based approaches in healthcare, as AI relies on the availability of data. The second category of ideas involves utilizing the model to investigate various ventilation maneuvers and settings, enabling the physicians to observe alterations in relevant quantities of interest, and decide for the best option. Moving to the highest level, there exists the possibility to use the model to suggest specific maneuvers or parameters for individual patients or even autonomously optimizing them in the long run.}

\textcolor{black}{Finally, we want to address the use of esophageal pressure measurements for the model calibration. While becoming increasingly common, those are not yet a standard practice in the medical management of ventilated patients. In this study, we utilized the measured esophageal pressure due to its inclusion in the example patient's study protocol, and our objective to also reproduce the volume-dependent pleural pressure with the presented lung model. In general, however, it is important to note that model calibration can be realized without direct measurement of the esophageal pressure, especially as long as all pressure-volume relations remain consistent throughout calibration and simulation. For example, in a previous study, we calibrated a simpler model without patient-specific esophageal pressure measurements~\cite{Roth2017a}. However, note that the model then operates on an artificial pleural pressure level, restricting its ability to represent actual lung stresses. As mentioned earlier, measured esophageal pressure qualitatively reflects pleural pressure behavior, although not quantitatively~\cite{Dornhorst1952,Cammarota2023}. Thus, using measured esophageal pressure might equally introduce a discrepancy in absolute values. Moreover, measurements in the esophagus are strongly influenced by various factors~\cite{Cammarota2023}, potentially leading to further inaccuracies. Despite these uncertainties, neither the parametrization nor the simulation results have been significantly compromised in this work. This indicates a certain robustness of the model to variations in pleural pressure. Further, we are often not only or mainly interested in absolute values, but rather in relative quantities, for instance for different ventilation maneuvers.}

\paragraph{Limitations}

The present model includes a few shortcomings along with the advantages and opportunities described.
To begin with, the optimization of input parameters for the described method to calibrate the model is improvable. 
Regarding the example patient presented, the RD time dependence is obviously not yet properly met by the chosen set of input parameters (e.g.,~see slope and amount of volume increase along the inspiratory hold maneuver in Figure~\ref{fig:results_insphold_occl0}). 
Therefore, we have to be cautious when interpreting the calibrated values of the critical pressures (Figure~\ref{fig:critpress}) and the absolute percentage of tidal RD currently given by the example model. So far, the RD dynamics in the model should only be evaluated qualitatively. 
We observed that the interaction of single input parameters and their influence on the resulting pressure-volume behavior \textcolor{black}{are} to some extent unpredictable. As a solution, the manual adaption of input parameters (Section~\ref{sec:mat&methods:parametrization:opti}) can be enhanced, e.g., by an inverse analysis used to automate the model calibration and obtain the best parameter fit by finding a global minimum for the optimization. In this context, we have already presented novel and very promising Bayesian based approaches that may be useful to identify the parameters for the present model~\cite{Hervas-Raluy2023,Willmann2022a,Nitzler2022}. Doing so enables both pure parameter optimization~\textcolor{black}{~---~maybe even for the full model without using the proposed calibration method to also integrate data from other measuring equipment~---~}as well as considering and testing different distributions of the time constants~$\tau$, e.g., quasi-hyperbolic, exponential, or lognormal, or the pathology-dependent design of variables like the time constants and tissue stiffnesses~$\kappa$~\cite{Roth2017a}. The latter goes beyond the scope of this study, but is a valid object of research as remodeling of tissue occurs already in an early stage of ARDS~\cite{Negri2002}.

\textcolor{black}{Furthermore, while the model introduces RD dynamics, it does not incorporate certain effects caused by the complex fluid mechanics at the air-liquid interface, that are known to appear during RD and that can affect airway walls and cells. These can be seen for example in numerous studies that investigate the complex injury processes leading to atelectrauma and the diverse characteristics that influence it~\cite{Novak2021}. In particular, the frequency of RD events and the presence of high gradients in wall pressure and shear stress appear to be detrimental~\cite{Bilek2003, Kay2004}, with the latter being influenced by various factors. There are interesting approaches in the literature to assess the damage caused by RD~\cite{Mellenthin2019, Jacob2012, Fujioka2016}. These and other approaches could provide valuable inspiration for extensions of the proposed model. Our current hypothesis is that these models and underlying insights could be used in an additive manner. One possibility would be to incorporate information on stresses, for instance from moving and rupturing liquid bridges, by adding it to the stress states directly obtained from the proposed model, once the model detects that recruitment occurs. Notably, the presented model might also provide quantities pivotal for determining the extent of RD-induced stress itself, such as opening velocities, or frequency and amount of RD. Naturally, further investigations are necessary to ensure a comprehensive reproduction of the injury potential.}

Another limitation in our model is the lack of local mechanical interdependence between the terminal units and of parenchymal tethering between tissue and conducting airway elements. We see an interaction between the terminal units through the coupling by a volume-dependent boundary condition of pleural pressure where the recruitment of some terminal units triggers a slight derecruitment in others, and vice versa. However, the model does not include the direct influence of neighbouring elements onto each other, especially onto their straining behavior when RD occurs. For the airway-tissue interaction, we consider local pressure differences by calculating the external pressure acting on an airway element as the
alveolar pressure of the closest terminal unit. We suppose that this already accounts for a large part of the airway-tissue interaction. However, the tethering of the lung parenchyma on the airway wall is neglected. There are promising approaches in literature to model those two types of interdependence~\cite{Ma2020,Roth2017,Ryans2019a}. Though, it is unclear whether the concepts are valid in the current context of RD dynamics and local pathologies, and how to couple them to the model. We therefore refrained from including them in the present work, and point to the need for further investigation of mechanical interdependence as it is assumed to be of importance regarding the phenomenon of RD~\cite{Broche2017,Mead1970}.

Lastly, we purely focused on the alveolar RD model and did not include airway collapse dynamics into the full lung model. Although the approach presented is able to mimic the key features of airway RD, the effect of gas trapping is neglected.
That shortcoming can easily be tackled by the additional application of the previously presented airway RD model~\cite{Geitner2022}. However, the problem of a proper parametrization of the critical RD pressures in the airways remains. Therefore, further research on the impact and also the clinical relevance of capturing gas trapping in the model may be advisable.

Despite the aforementioned limitations, the newly introduced alveolar RD components and their integration into a comprehensive multi-compartment model of the lung offer promising and relevant opportunities, as outlined above. 
In conclusion, the approach presented herein has the potential to individually estimate the benefit or risk of VILI caused by arbitrary ventilation strategies, and to significantly contribute to improving clinical ventilation therapy.

\clearpage
\appendix
\label{sec:appendix}

\section{Conducting airways}
\label{sec:mat&methods:airways}

Each airway branch used in this study is modelled by a reduced-dimensional airway element that mimics the averaged behavior of flow and wall mechanics of a fully resolved, elastic three-dimensional airway~\cite{Ismail2013}. The pressure drop $\Delta P = P_{\mathrm{in}} - P_{\mathrm{out}}$ across a 0D airway element and the external pressure~$\widetilde{P}_{\mathrm{ext}}$ cause the inflow~$Q_{\mathrm{in}}$ and outflow~$Q_{\mathrm{out}}$ of the element which can be calculated according to
%
\begin{equation}
	\begin{aligned}
	C \dfrac{\mathrm{d}}{\mathrm{d}t}\left( \dfrac{1}{2} \left( P_{\mathrm{in}} + P_{\mathrm{out}} \right) - \widetilde{P}_{\mathrm{ext}}\right)  + Q_{\mathrm{out}} - Q_{\mathrm{in}} = 0,\\
\dfrac{I}{2} \dfrac{\mathrm{d}}{\mathrm{d}t} \left(Q_{\mathrm{in}} + Q_{\mathrm{out}}\right) + R \cdot \left( Q_{\mathrm{in}} + Q_{\mathrm{out}} \right) + P_{\mathrm{out}} - P_{\mathrm{in}} = 0,
	\end{aligned}
	\label{eq:0D_pipe}
\end{equation}
where $C$ is the capacitance of the airway wall, $R$ a generation-dependent airway resistances, and $I$ the inductance. For further details on these quantities \textcolor{black}{see our previous work~\cite{Roth2017a,Ismail2013,Roth2017}}.

\section{Parameters of patient-specific lung model}
\label{sec:appendix_parameters}

\subsection{Pressure boundary condition}

Table~\ref{tab:params_ppl} specifies the parameters used for the volume-dependent pleural pressure component $P_{\mathrm{pl}}^\mathrm{vol}$ and the static contribution~$P_{\mathrm{pl}}^\mathrm{weight}$, both introduced in Section~\ref{sec:mat&methods:pressBC} 
and fit from clinical data of the example patient.

\begin{table}[ht]
\centering
\begin{tabular}{l|c|c} 
 \hline
 Parameter & Value & Units \\ [0.5ex] 
 \hline\hline
 $a_\mathrm{v}$ & 5.3 & mbar \\
 $b_\mathrm{v}$ & 3.0 & mbar \\
 $c_\mathrm{v}$ & $2.3 \cdot 10^{-1}$ & mbar \\
 $d_\mathrm{v}$ & 2.0 & [-] \\
 \hline
 $a_\mathrm{w}$ & 14.3 & mbar \\
 $b_\mathrm{w}$ & $2.5 \cdot 10^{-1}$ & $\mathrm{mbar.mm}^{-1}$\\
 $c_\mathrm{w}$ & $1.6 \cdot 10^{-3}$ & $\mathrm{mbar.mm}^{-2}$\\
 $d_\mathrm{w}$ & $5.1 \cdot 10^{-6}$ & $\mathrm{mbar.mm}^{-3}$\\
 $e_\mathrm{w}$ & $6.2 \cdot 10^{-9}$ & $\mathrm{mbar.mm}^{-4}$\\
 \hline
\end{tabular}
\caption{Parameters specifying the pleural pressure $P_{\mathrm{pl}}$ (Section~\ref{sec:mat&methods:pressBC}) for the example patient.}
\label{tab:params_ppl}
\end{table}

\subsection{Calibration of terminal units}

The parameters entering the calibration of the terminal units in the lung model of our example patient (Section~\ref{sec:mat&methods:parametrization}) are listed in the following tables. Table~\ref{tab:input_params} provides the input parameters and Table~\ref{tab:vent_params} lists the ventilator setting and ventilation measurements used for the parametrization procedure described. 

\begin{table}[ht]
\centering
\begin{tabular}{l|c|c} 
 \hline
 Parameter & Value & Units\\ [0.5ex] 
 \hline\hline
 $\kappa$ & 13.0 & mbar \\ 
 $\beta$ & -8.0 & [-]\\
 $\Delta P_{\mathrm{max-min}}$ & 12.0 & mbar\\
 $\Delta P_{\mathrm{op-cl}}$ & 4.0 & mbar\\
 $T$ & 5.0 & s\\
 $\mu_{\mathrm{cl}}$ & -2.5 & mbar\\
 $\sigma_{\mathrm{cl}}$ & 12.0 & mbar\\
 $\mu_{\mathrm{op}}$ & 9.2 & mbar\\ 
 $\sigma_{\mathrm{op}}$ & 50.0 & mbar\\ 
 $k_{\mathrm{coll}}$ & $2.9 \cdot 10^{-2} $ & [-]\\ 
 $k_{\mathrm{edema}}$ & 3.0 & [-]\\
 $\epsilon_\mathrm{V_\mathrm{0}}$ & $1.4 \cdot 10^{-3} $ & [-]\\
 \hline
\end{tabular}
\caption{Input parameters for RD parametrization (Section~\ref{sec:mat&methods:parametrization}) fitted for the example patient.}
\label{tab:input_params}
\end{table}
\begin{table}[ht]
\centering
\begin{tabular}{l|c|c} 
 \hline
 \multicolumn{3}{c}{Ventilation parameters}  \\ [0.5ex] 
 \hline
 Parameter & Value & Units \\
 \hline\hline
 PEEP & 8.0 & mbar\\ 
 $P_{\mathrm{endinsp}}$ & 20.0 & mbar\\
 $P_{\mathrm{tp,PEEP}}$ \textit{(global)} & 2.5 & mbar\\
 $P_{\mathrm{tp,endinsp}}$ \textit{(global)} & 11.5 & mbar\\
 $t_{\mathrm{insp}}$ & 1.2 & s\\
 $t_{\mathrm{exp}}$ & 2.1 & s\\ 
 $P_{\mathrm{tp,qs,start}}$ & 1.3 & mbar\\
 $P_{\mathrm{tp,qs,max}}$ & 12.0 & mbar\\
 $t_{\mathrm{insp,qs}}$ & 14.9 & s\\
 $t_{\mathrm{exp,qs}}$ & 15.8 & s\\
 \hline
\end{tabular}
\caption{Ventilation parameters of the example patient used for RD parametrization (Section~\ref{sec:mat&methods:parametrization}).}
\label{tab:vent_params}
\end{table}

\clearpage
\subsection{Additional simulation results}

\begin{figure}[ht]
    \centering
    \includegraphics[page=1,trim=0 0 0 0,clip,width=0.7\textwidth]{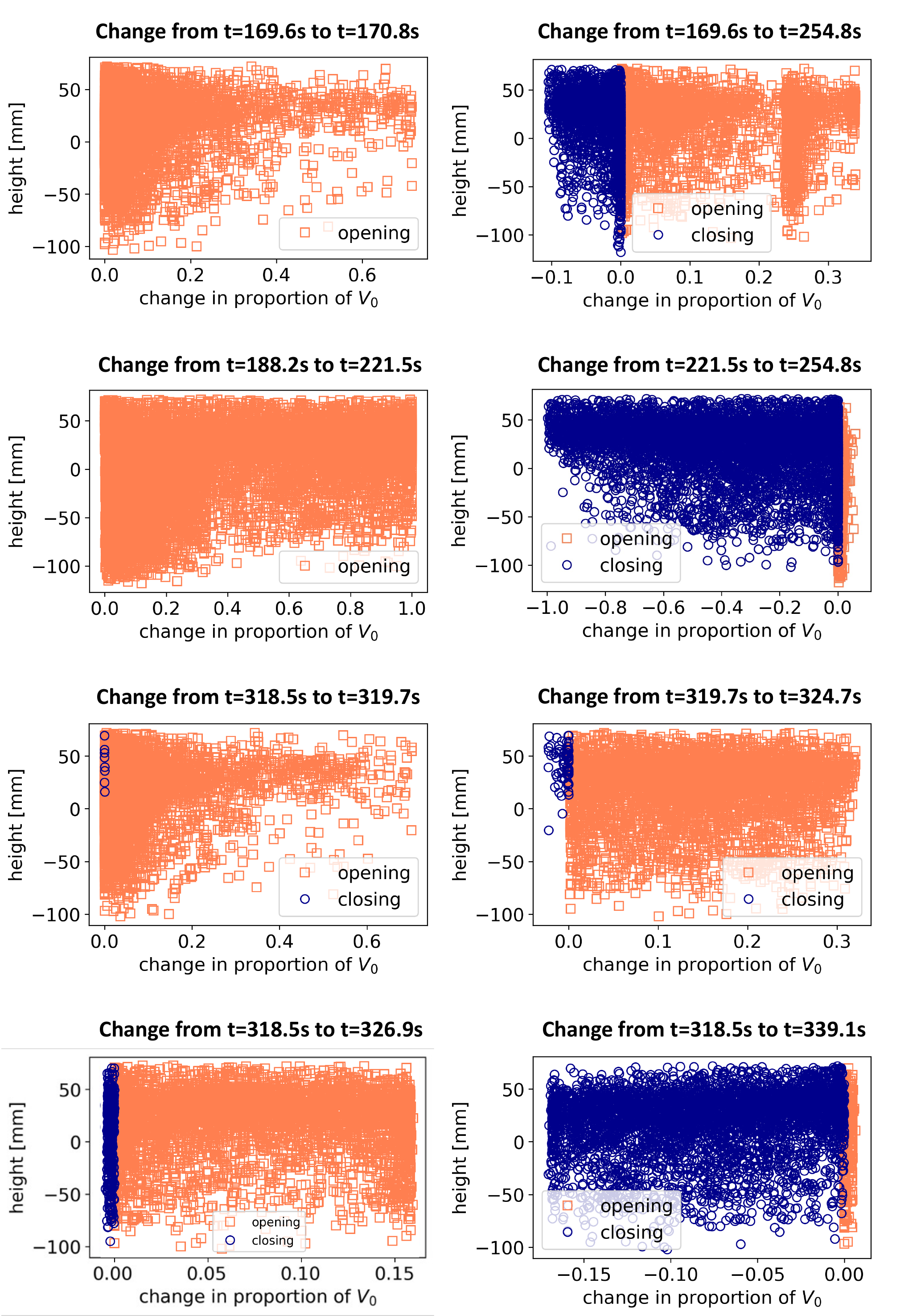}
    \caption{Proportional change in the reference volume of all terminal units affected by RD between two specified time points; orange squares indicate recruiting and blue circles derecruiting terminal unit, respectively; the coordinates of the lung height range from approximately 70~mm (ventral) to -110~mm (dorsal).}
    \label{fig:results_local_RD}
\end{figure}
%


\section*{Acknowledgements}

We gratefully acknowledge financial support by the Deutsche Forschungsgemeinschaft (DFG, German Research Foundation) in the project WA1521/26-1, and by BREATHE, a Horizon 2020-ERC-2020-ADG project (grant agreement No. 101021526-BREATHE).

\section*{Disclosures}

No conflicts of interest, financial or otherwise, are declared by the authors.


\bibliography{paper.bib}

\printnomenclature

\end{document}